\RequirePackage{lineno}
\documentclass[twocolumn,showpacs,superscriptaddress,preprintnumbers,amsmath,amssymb,prc]{revtex4}  
\usepackage{graphicx}
\usepackage{dcolumn}
\usepackage{bm}
\usepackage{float}
\usepackage{color}
\usepackage{CJK}
\usepackage[colorlinks,linkcolor=blue,urlcolor=blue,citecolor=blue]{hyperref}
\usepackage{isotope}
\usepackage[normalem]{ulem}
\renewcommand{\sout}{\bgroup \color{red} \ULdepth=-0.5ex \ULset}

\usepackage{amsmath}
\usepackage{cases}
\usepackage{amssymb}
\usepackage{eqnalign}

\begin{document}
\begin{CJK*}{UTF8}{gbsn}
	
\title{Dynamical development of proton cumulants and correlation functions in Au+Au collisions at $\sqrt{s_{NN}} = 7.7$ GeV from a multiphase transport model}
\author{Qian Chen}
\affiliation{Shanghai Institute of Applied Physics, Chinese Academy of Sciences, Shanghai 201800, China}
\affiliation{University of Chinese Academy of Sciences, Beijing 100049, China}
\affiliation{Key Laboratory of Nuclear Physics and Ion-beam Application (MOE), Institute of Modern Physics, Fudan University, Shanghai 200433, China}
\affiliation{Shanghai Research Center for Theoretical Nuclear Physics, NSFC and Fudan University, Shanghai $200438$, China}	
\author{Guo-Liang Ma}
\email[]{glma@fudan.edu.cn}
\affiliation{Key Laboratory of Nuclear Physics and Ion-beam Application (MOE), Institute of Modern Physics, Fudan University, Shanghai 200433, China}
\affiliation{Shanghai Research Center for Theoretical Nuclear Physics, NSFC and Fudan University, Shanghai $200438$, China}
\affiliation{Shanghai Institute of Applied Physics, Chinese Academy of Sciences, Shanghai 201800, China}
	
	
\begin{abstract}
Higher-order cumulants of the distributions of conserved charges, such as net-baryon number, are sensitive to the quantum chromodynamics(QCD) phase transition and the QCD critical point. We calculate the cumulants and correlation functions of proton, antiproton, and net-proton multiplicity distributions in Au+Au collisions at $\sqrt{s_{NN}} = 7.7$ GeV using a multiphase transport model(AMPT). The AMPT model can basically describe the trends of cumulants, cumulant ratios, (normalized) correlation functions of the proton and net-proton measured by the STAR experiment. The multiproton (baryon) correlations in the AMPT model are consistent with the expectation from baryon number conservation. We demonstrate that multiproton (baryon) correlations suffer the dynamical evolution of heavy-ion collisions. Our results provide a baseline for searching for the possible critical behaviors at the critical end point in relativistic heavy-ion collisions.
\end{abstract}
	
\maketitle
	
\section{Introduction}
\label{framework}

\label{introduction}
Exploring the phase structure of quantum chromodynamics (QCD) is one of the important frontiers in both theoretical and experimental researches of strongly interacting QCD matter in the past decade. Theoretically, the QCD phase diagram has been extensively studied using lattice QCD from first principles~\cite{Ding:2015ona,HotQCD:2014kol,Bellwied:2013cta,Borsanyi:2014ewa}, functional renormalization group methods~\cite{Zhang:2017icm,Mitter:2014wpa,Herbst:2013ufa,Fu:2019hdw}, and effective models~\cite{Fukushima:2003fw,Skokov:2010uh,Pisarski:2016ixt,Li:2017ple}. Experimentally, the phase structure of strongly interacting QCD matter can be accessed by relativistic heavy-ion collisions~\cite{Bzdak:2019pkr,Luo:2020pef}.

The lattice QCD results have shown that the phase transition at a vanishing baryon chemical potential ($\mu _{B}$)  is a cross over~\cite{Aoki:2006we} with a pseudocritical temperature of $T_{C} \simeq 160$ MeV~\cite{Aoki:2009sc,Bazavov:2011nk}, but not any constraint on the phase transition at large $\mu _{B}$ is currently available. While in the large $\mu _{B}$ region, the phase transition could be the first order~\cite{Ejiri:2008xt}. The end point of the possible first-order phase boundary toward the crossover region is called the QCD critical end point (CEP)~\cite{Skokov:2010uh,Aoki:2006we,Karsch:2011gg}, as an analog of the critical point in the phase diagram of water.  The divergence of correlation length occurs at the CEP near the second-order transition~\cite{Jeon:2000wg,Koch:2008ia,Bzdak:2018uhv}, while the first order phase transition leads to possible droplet formation~\cite{Mishustin:1998eq,Xu:2007oam,Jin:2018fwq} or spinodal instabilities~\cite{Chomaz:2003dz,Randrup:2003mu,Sasaki:2007db,Steinheimer:2012gc}. The location of the CEP and the first-order coexistence region has not really been determined due to theoretical limitations~\cite{Wu:2021xgu}, so supplementary experimental effort is needed to explore the QCD phase structure.

In order to map the QCD phase diagram, extensive experimental work has been carried out in relativistic heavy ion collision experiments at the BNL Relativistic Heavy Ion Collider (RHIC),  the CERN Super Proton Synchrotron (SPS), and CERN Large Hadron Collider (LHC) ~\cite{STAR:2005gfr,STAR:2010mib,BRAHMS:2004adc,Friman:2011pf}. There will be further devotions from the future accelerator facilities, including the Facility for Antiproton and Ion Research (FAIR),  the Nuclotron-based Ion Collider Facility (NICA), and the High Intensity Heavy-ion Accelerator Facility (HIAF). Because both the CEP and the first-order phase transition are associated with the characteristics of fluctuations and correlations, many experimental observables have been proposed to capture these features. The cumulants of conserved charge distribution have received lots of attention recently, because they are expected to be very sensitive to the correlation length~\cite{Stephanov:2008qz,Athanasiou:2010kw,Stephanov:2011pb,Cheng:2008zh,Gavai:2010zn,Chen:2021kjd} and can be directly related to the susceptibilities in lattice QCD calculations~\cite{Gupta:2011wh,Ding:2015ona,Bazavov:2012vg}.

For a system in thermal equilibrium, a grand-canonical ensemble can be characterized by its dimensionless pressure, which is the logarithm of the QCD partition function~\cite{Ding:2015ona},
\begin{eqnarray}
\frac{P}{T}=\frac{1}{VT^{3}}\ln\left [ Z\left ( V,T,\mu _{B},\mu _{Q},\mu _{S} \right ) \right ], \label{MDIV}
\end{eqnarray}%
where $V$ and $T$ are the system volume and temperature, the $\mu _{B}$, $\mu _{Q}$, and $\mu _{S}$ are baryon, charge, and strangeness chemical potential, respectively. For a thermodynamical system with different degrees of freedom and interactions, it can be identified by fluctuations of conserved charges. In lattice QCD, these fluctuations can be quantified in terms of the susceptibilities of conserved charges
\begin{eqnarray}
\chi _{ijk}^{BQS}=\frac{\partial^{(i+j+k)}\left [ P/T^{4} \right ] }{\partial\hat{\mu _{B}^{i}}\partial\hat{\mu _{Q}^{j}}\partial\hat{\mu _{S}^{k}}},\label{MDIV1}	
\end{eqnarray}%
where $\hat{\mu _{q}}=\mu _{q}/T$, $q = B, Q, S$. The corresponding cumulants can be obtained from the susceptibilities of these conserved charges~\cite{Ding:2015ona},
\begin{eqnarray}
C_{ijk}^{BQS}&=&\frac{\partial ^{(i+j+k))}\ln[Z(V,T,\mu _{B},\mu _{Q},\mu _{S})]}{\partial\hat{\mu _{B}^{i}}\partial\hat{\mu _{Q}^{j}}\partial\hat{\mu _{S}^{k}}} \notag \\
&=&VT^{3}\chi _{ijk}^{BQS}(T,\mu _{B},\mu _{Q},\mu _{S}).\label{MDIV2}
\end{eqnarray}%
From Eqs.~(\ref{MDIV1}) and (\ref{MDIV2}), it is convenient for us to consider the ratios of cumulants, because they are intensive and volume-independent in the thermodynamic limit. The cumulant ratios are defined as follows~\cite{Ding:2015ona}:
\begin{eqnarray}
\frac{\chi _{2}}{\chi _{1}}=\frac{C_{2}}{C_{1}}=\frac{\sigma ^{2}}{M},\frac{\chi _{3}}{\chi _{2}}=\frac{C_{3}}{C_{2}}=S\sigma ,\frac{\chi _{4}}{\chi _{2}}=\frac{C_{4}}{C_{2}}=\kappa \sigma ^{2},\label{MDIV3}
\end{eqnarray}%
where the cumulants of event-by-event conserved charge multiplicity distribution, which are represented by $C_{n}$,  are linked to mean ($M$), variance ($\sigma ^{2}$), skewness ($S$), kurtosis ($\kappa$).  

The RHIC has been performing a beam energy scan program since 2010~\cite{STAR:2010vob,STAR:2017sal,STAR:2010mib,STAR:2014egu,STAR:2013gus}. The RHIC-STAR experiment has measured the cumulants (up to the sixth order) of net-proton (proxy of net-baryon), net-charge, and net-kaon multiplicity distributions in Au+Au collisions at $\sqrt{s_{NN}} = 7.7$, 11.5, 14.5, 19.6, 27, 39, 54.4, 62.4, and 200 GeV~\cite{STAR:2014egu,STAR:2013gus,Pandav:2020uzx,Thader:2016gpa}, which correspond to a baryon chemical potentials range from 420 to 20 MeV. Among these conserved charges, there are some studies suggesting that the cumulants of the baryon number are most sensitive to the phase transition of QCD matter~\cite{Asakawa:2000wh, Asakawa:2009aj}. The recent results from the STAR Collaboration for the ratio of fourth-order over second-order net-proton cumulants ($\kappa \sigma ^{2}$) show a nonmonotonic energy dependence with a significance of 3.1$\sigma$~\cite{STAR:2021iop}. The observed large four-particle correlation could be attributed to the formation of proton clusters and related to the characteristics of CEP or first-order phase transition~\cite{Bzdak:2016sxg,Ling:2015yau,Fu:2021oaw}.

In this paper, we focus on the dynamical evolution of the cumulants and multiparticle correlation of proton and baryon multiplicity distributions using a multiphase transport model(AMPT), because these fluctuations and correlations have to suffer dynamical evolution stages of heavy-ion collisions.  It is noticeable that the dynamical descriptions of proton and baryon cumulants and correlation functions for central Au+Au collisions at $\sqrt{s_{NN}} = 7.7$--200 GeV have been recently investigated by viscous hydrodynamic simulations~\cite{Vovchenko:2021kxx}. Our paper is organized as follows. In Sec. II, we briefly introduce the AMPT model and how we calculate cumulants and correlation functions. In Sec. III, we compare our model results with the measurements from the STAR experiment, and discuss the evolution of the cumulants and multiparticle correlation of proton and baryon. In Sec. IV, a summary is provided.

\section{Model and calculation method}
\label{framework}

\subsection{The AMPT model}
\label{ampt}
A multiphase transport model AMPT~\cite{Lin:2004en,Lin:2021mdn} is widely used to study the physics of relativistic heavy-ion collisions ~\cite{Ma:2016fve,Ma:2011uma,Ma:2013pha,Bozek:2015swa,Bzdak:2014dia}. The AMPT model with string meting mechanism consists of four main components: initial condition, parton cascade, hadronization, and hadronic rescatterings. The initial condition mainly provides the spatial and momentum distributions of minijet partons from QCD hard processes and soft string excitations by using
the HIJING model~\cite{Wang:1991hta,Gyulassy:1994ew}. The parton cascade describes the partonic evolution with a quark-antiquark plasma resulting from the melting of excited strings and minijets. Parton scatterings are modeled by Zhang's parton cascade (ZPC)~\cite{Zhang:1997ej}, which currently only includes two-body elastic parton scatterings using perturbative QCD cross sections (3 mb) with a screening mass. When all partons stop to interact, a simple quark coalescence model is then used to combine nearby partons into hadrons. The interactions of baryon-baryon, baryon-meson, and meson-meson in hadronic phase are described by a relativistic transport (ART) model~\cite{Li:1995pra}. We also include resonance decays, including those of unstable strange hadrons (e.g., $\Lambda$, $\Sigma$, $\Xi$, and $\Omega$) which provide a feed-down contribution to protons and antiprotons. 

In this work, a new version of the AMPT model, which ensures the conservation of various charges (charge, baryon number, and strangeness number) in each hadronic reaction channel during hadronic evolution, is used to study the fluctuations and correlations of baryon and proton numbers. There are two main reasons why the total charge is not conserved in the old version of  AMPT model~\cite{Lin:2014uwa}. 1) In the old version, only $K^{+}$ and $K^{-}$ were introduced in hadron rescatterings as explicit particles, but $K^{0}$ and $\bar{K^{0}}$ were omitted. In order to include $K^{0}$ and $\bar{K^{0}}$ effectively, the old version replaced $K^{0}$ with $K^{+}$ and $\bar{K^{0}}$ with $K^{-}$ before hadron rescatterings. And the model replaced half of $K^{+}$ with $K^{0}$ and half of $K^{-}$ with $\bar{K^{0}}$ after hadron rescatterings. This leads to a certain degree of the violation of total charge conservation. 2) In the old version, not all possible isospin configurations were considered for hadronic reaction channels or resonance decays. The isospin-averaged cross sections were used instead, and the charge of the final state particles is chosen randomly from all possible charges, independent of the total charge of the initial state. It made the total charge not be conserved. For example,  $\pi^{+}$+$\pi^{+}$ should be allowed to enter $\rho^{+}$+$\rho^{+}$ instead of $\rho^{+}$+$\rho^{-}$ and $\rho^{-}$+$\rho^{-}$. To solve these two main problems, $K^{0}$ and $\bar{K^{0}}$ have been introduced as explicit particles in the charge conservation version of the AMPT model. On the other hand, all problematic reaction channels have been corrected to ensure that all reaction channels satisfy the conservation of electric charge, baryon number, and strangeness number. The principle of detailed balance have also been implemented to ensure that each reaction channel is in equilibrium with its opposite channel at equilibrium. The charge conservation version of AMPT model has shown shown good performance in describing the STAR measured net-charge fluctuations~\cite{Huang:2021ihy}. It should be important for studying baryon number fluctuations to correctly model all reaction channels, especially meson-baryon and baryon-baryon scatterings. For example, $\pi N$ reactions are considered to be responsible for randomizing nucleon isospin during hadronic phase evolution~\cite{Kitazawa:2011wh,Kitazawa:2012at}. In this work, we use the charge conservation version of AMPT model to collect a total of 70 million events to study cumulants and correlations in Au+Au collisions at $\sqrt{s_{NN}} = 7.7$ GeV.

\subsection{Calculation method}
\label{sec:partB}
In statistics, various features of the probability distribution can be characterized by different moments, such as mean ($M$), variance ($\sigma ^{2}$), skewness ($S$), kurtosis ($\kappa$). For example, the event-by-event particle multiplicity distribution can be characterized by the cumulants of multiplicity distribution, from which correlation functions can be obtained. The various orders of cumulants $C_{n}$ of multiplicity distribution can be calculated as follows~\cite{Luo:2011rg,Luo:2010by,Luo:2011ts,Luo:2017faz}:
\begin{eqnarray}
C_{1}&=&\left \langle N \right \rangle,\notag \\
C_{2}&=&\left \langle \left (\delta N  \right )^{2} \right \rangle,\notag \\
C_{3}&=&\left \langle \left (\delta N  \right )^{3} \right \rangle,\notag \\
C_{4}&=&\left \langle \left (\delta N  \right )^{4} \right \rangle - 3\left \langle \left (\delta N  \right )^{2} \right \rangle^{2},\label{MDIV}
\end{eqnarray}%
where $N$ and $\bar{N}$ are the numbers of particles and antiparticles on an event-by-event bias, $\delta N=N-\bar{N}$ and $\left \langle \cdots \right \rangle$ represents an event average. With the definition of cumulants, various moments can be obtained as follows:
\begin{eqnarray}
M=C_{1},\sigma ^{2}=C_{2},S=\frac{C_{3}}{\left ( C_{2} \right )^{\frac{3}{2}}},\kappa =\frac{C_{4}}{(C_{2})^{2}}.
\label{MDIV}	
\end{eqnarray}%
In addition, the moment products can be expressed in terms of the ratios of cumulants as shown in Eq.~(\ref{MDIV3}).

However, cumulants have a disadvantage because they mix different orders of correlations, so it is more instructive to study (integrated) multiparticle correlation functions~\cite{Bzdak:2019pkr,Bzdak:2016sxg,Kitazawa:2017ljq}. The following relations can be used to access the integrated $n$-particle correlation functions $\kappa _{n}$ (also known as factorial cumulants):
\begin{eqnarray}
\kappa _{1}&=&C_{1}=\left \langle N \right \rangle,\notag \\
\kappa _{2}&=&-C_{1}+C_{2},                         \notag \\
\kappa _{3}&=&2C_{1}-3C_{2}+C_{3},                   \notag \\
\kappa _{4}&=&-6C_{1}+11C_{2}-6C_{3}+C_{4},           \label{MDIV0}
\end{eqnarray}%
and vice versa, i.e.,
\begin{eqnarray}
C_{2}&=&\kappa _{2}+\kappa _{1},                          \notag \\
C_{3}&=&\kappa _{3}+3\kappa _{2}+\kappa _{1},              \notag \\
C_{4}&=&\kappa _{4}+6\kappa _{3}+7\kappa _{2}+\kappa _{1}.    \label{MDIV4}
\end{eqnarray}%
Meanwhile, the cumulant ratios can be expressed in terms of normalized correlation functions ${\kappa _{n}}/{\kappa _{1}}$ ($n >$ 1) as
\begin{eqnarray}
\frac{C_{2}}{C_{1}}&=&\frac{\kappa _{2}}{\kappa _{1}}+1,         \notag \\
\frac{C_{3}}{C_{2}}&=&\frac{\kappa _{3}/\kappa _{1}-2}{\kappa _{2}/\kappa_{1}+1}+3,   \notag \\
\frac{C_{4}}{C_{2}}&=&\frac{\kappa _{4}/\kappa _{1}+6\kappa _{3}/\kappa _{1}-6}{\kappa _{2}/\kappa_{1}+1}+7.\label{MDIV5}
\end{eqnarray}%
 
In our calculations, we apply the same kinematic cuts as used in the STAR experimental analysis~\cite{STAR:2021iop,STAR:2020tga} to calculate the aforementioned different cumulants of proton and antiproton multiplicity distributions. We select protons and antiprotons within a transverse momentum range of $0.4<p_{T}<2.0$ GeV/$c$ and a midrapidity window of $\left |y   \right |< 0.5$. We use the charged particle multiplicity distribution to define centrality bins. In order to avoid the self-correlation, the charged particle multiplicity other than protons and antiprotons within pseudorapidity $\left |\eta   \right |< 1$ is used. We apply the formulas based on the $\Delta$ theorem to estimate the statistical error calculations. For more detailed information on error calculations, please refer to Refs.~\cite{Luo:2017faz,STAR:2021iop}.

\subsection{Baryon number conservation}
\label{sec:partC}

On the other hand, the total baryon number is absolutely conserved in relativistic heavy-ion collisions, which is defined by the total nucleon number coming from projectile and target nuclei. The baryon number conservation is also satisfied in the AMPT model. In the initial state, the baryon number is decelerated to the midrapidity zone due to the baryon stopping. In a proposed baryon stopping case~\cite{Bzdak:2019pkr,Bzdak:2016jxo}, if the distribution of the number of measured baryons $N$ can be simply given by a binomial distribution,
\begin{eqnarray}
P\left ( N \right )=\frac{B!}{N!(B-N)!}p^{N}\left ( 1-p \right )^{(B-N)},\label{MDIV}
\end{eqnarray}%
where $p$ is the probability that an initial nucleon will eventually stop and enter the acceptance window, and $B$ is the total number of baryons, the induced $n$-baryon correlation $\kappa _{n}$ holds as follows:
\begin{eqnarray}
\kappa _{1}=\left \langle N \right \rangle=pB, \kappa _{2}=-\frac{\left \langle N \right \rangle^{2}}{B}, \kappa _{3}=2\frac{\left \langle N \right \rangle^{3}}{B^{2}}, \kappa _{4}=-6\frac{\left \langle N \right \rangle^{4}}{B^{3}}. \label{corrbnc}	
\end{eqnarray}%
It suggests that baryon number conservation leads to a $n$-baryon correlation $\kappa _{n}$ with the sign of $(-1)^{n+1}$ and the strength proportional to $\left \langle N \right \rangle ^{n}$.

\section{Results and Discussions}
\label{framework}

\subsection{Centrality bin width effect and correction}
In order to compare with experimental results from the STAR collaboration, the AMPT events are divided into nine collision centrality classes. For the two bins with largest number of participant nucleons $ N_{part} $, they correspond to 0--5 \% and 5--10 \% of most central collisions, while the other seven bins are 10 \% wide covering the remaining 10--80 \% of most central collisions. However, it could induce a centrality bin width effect (CBWE) to calculate cumulants or cumulant ratios in a broad centrality bin, because the CBWE can be caused by volume fluctuations in a wide centrality bin~\cite{Luo:2011ts,Luo:2013bmi,He:2018mri,Huang:2021ihy}. We know that collision centrality classes are determined by the distribution of particle multiplicity, which is actually the smallest centrality bin. In other words, if the cumulants are calculated in each multiplicity bin and weighted by the number of events in each bin, we can eliminate the influence of the CBWE. This is the STAR proposed centrality bin width  correction(CBWC)~\cite{Luo:2017faz,STAR:2021iop}, which can be expressed as
\begin{eqnarray}
C_{n}=\frac{\sum_{r}n_{r}C_{n}^{r}}{\sum_{r}n_{r}},\label{MDIV}
\end{eqnarray}%
where $C_{n}$ represents any cumulants or cumulant ratio for one centrality bin, $n_{r}$ is the number of events in the $r$th multiplicity bin, $C_{n}^{r}$ represents the cumulant or cumulant ratio in the $r$th multiplicity bin.
\begin{figure}[htb]
\centering
\includegraphics
[width=9cm]{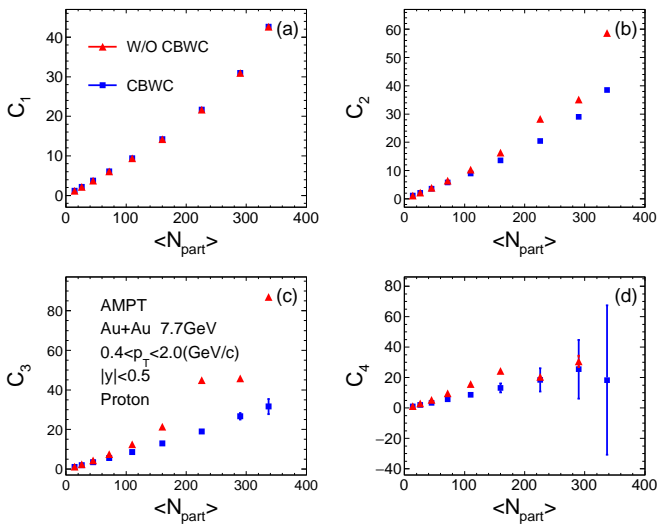}
\caption{(Color online)  The AMPT results on cumulants $C_{n}$ ($n=1$, 2, 3, and 4) of proton distributions as a function of $\left \langle N_{part} \right \rangle$ in Au+Au collisions at $\sqrt{s_{NN}} = 7.7$ GeV.}
\label{FIG.1.}
\end{figure}

We calculate the cumulants with and without the CBWC to see the impact of the CBWE. Figure~\ref{FIG.1.} shows the cumulants $C_{n}$ of proton distributions up to the fourth order as a function of $\left \langle N_{part} \right \rangle$ in Au+Au collisions at $\sqrt{s_{NN}} = 7.7$ GeV. In Fig.~\ref{FIG.1.}(a), the first order of cumulant $C_{1}$ without the CBWC is consistent with the CBWC, indicating that the influence of CBWE on the mean $M$ of the proton distribution is negligible. For $C_{n}$ ($n=2$, 3, and 4), the results without the CBWC are significantly greater than those with the CBWC. We find that the influence of CBWE increases with the increasing of $\left \langle N_{part} \right \rangle$. Note that the value of $C_{4}$ without the CBWC is $-$500.31$\pm $8.33 for the most central centrality bin, which is not shown in Fig.~\ref{FIG.1.}(d). It indicates that the CBWE contributes significantly to higher order cumulants, especially for more central collisions. Therefore, the CBWC is very essential to suppress the effect of the volume fluctuations on cumulants within a finite width of centrality bin.  We apply the CBWC to all cumulants and cumulant ratios in our calculations. For comparison with STAR data,  the $\Delta$ theorem method is applied to calculate the statistical errors with considering the CBWC~\cite{STAR:2021iop}.

\subsection{Centrality dependence}
\label{sec:partC}
The STAR Collaboration has recently published the centrality and energy dependences of the cumulants and correlation functions of proton, antiproton and net-proton multiplicity distributions for Au+Au collisions at $\sqrt{s_{NN}} = 7.7$, 11.5, 14.5, 19.6, 27, 39, 54.4, 62.4, and 200 GeV~\cite{STAR:2021iop}. In this subsection, we focus on the centrality dependence of cumulants, cumulant ratios, correlation functions, and normalized correlation functions of proton, antiproton, and net-proton in Au+Au collisions at $\sqrt{s_{NN}} = 7.7$ GeV. We compare our results with the published STAR data. 
\begin{figure}[htb]
\centering
\includegraphics
[width=9cm]{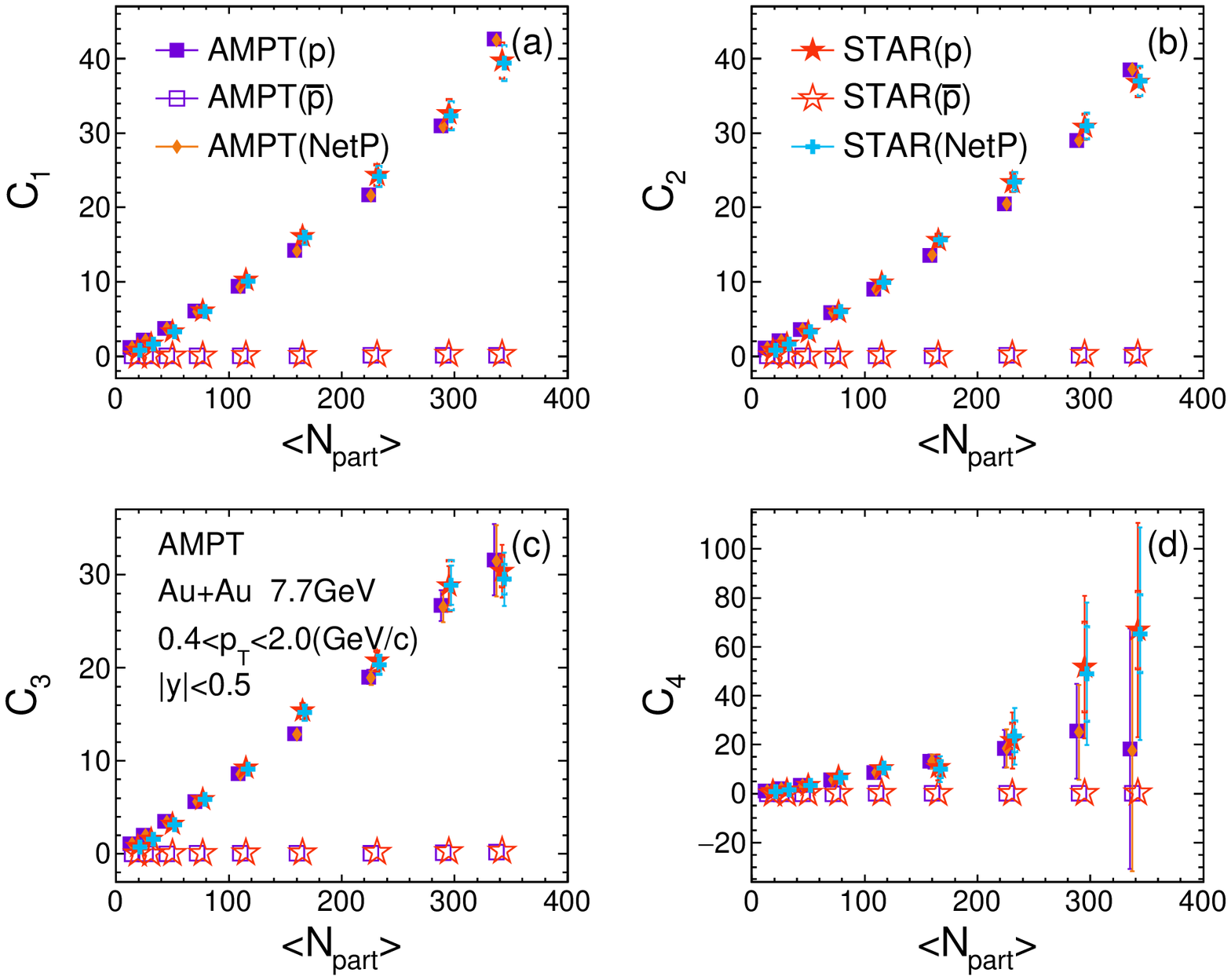}
\caption{(Color online) The AMPT results on cumulants $C_{n}$ of proton, antiproton, and net-proton distributions as a function of $\left \langle N_{part} \right \rangle$ in Au+Au collisions at $\sqrt{s_{NN}} = 7.7$ GeV, in comparisons with the STAR measurements~\cite{STAR:2021iop}.}
\label{FIG.2.}
\end{figure}
\begin{figure*}[htb]
\centering
\includegraphics
[width=17.5cm]{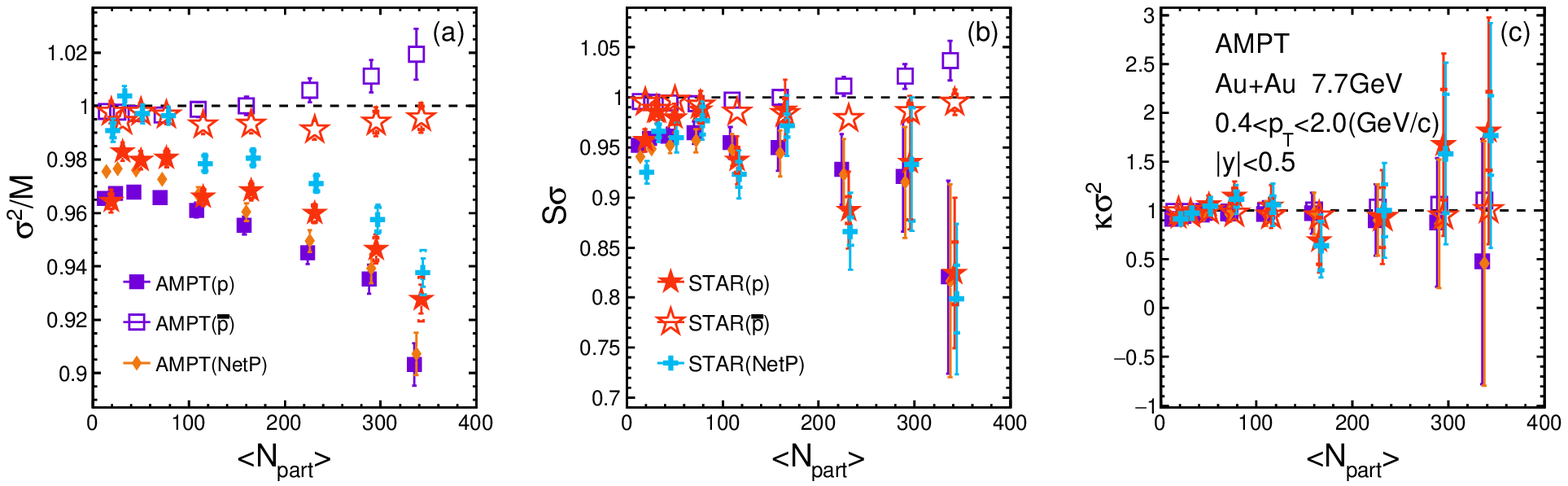}
\caption{(Color online) The AMPT results on cumulant ratios of proton, antiproton, and net-proton distributions as a function of $\left \langle N_{part} \right \rangle$ in Au+Au collisions at $\sqrt{s_{NN}} = 7.7$ GeV, in comparisons with the STAR measurements~\cite{STAR:2021iop}.}
\label{FIG.3.}
\end{figure*}

Figure~\ref{FIG.2.} shows the $\left \langle N_{part} \right \rangle$ dependence of cumulants $C_{n}$ ($n=1$, 2, 3, and 4) of proton, antiproton, and net-proton multiplicity distributions in Au+Au collisions at $\sqrt{s_{NN}} = 7.7$ GeV. As $\left \langle N_{part} \right \rangle$ increases, the cumulants $C_{n}$ ($n=1$, 2, 3, and 4) of the antiproton are almost zero because of the few yields of the antiproton. Therefore, the net-proton $C_{n}$ are mainly from protons. The cumulants $C_{n}$ of proton and net-proton all exhibit an increasing dependence of $\left \langle N_{part} \right \rangle$. We find that the AMPT model can basically describe the experimental data.

Figure~\ref{FIG.3.} shows the $\left \langle N_{part} \right \rangle$ dependence of cumulant ratios of proton, antiproton, and net-proton multiplicity distributions [$C_{2}/C_{1}\left ( \sigma ^{2}/M \right )$, $C_{3}/C_{2}\left ( S\sigma \right )$, and $C_{4}/C_{2}\left ( \kappa \sigma^{2} \right )$] in Au+Au collisions at $\sqrt{s_{NN}} = 7.7$ GeV, because these cumulant ratios are expected to eliminate the possible volume effect. In Fig.~\ref{FIG.3.}(a), the AMPT results show a consistent trend with experimental measurements of $\sigma ^{2}/M$, but with a slightly smaller magnitude. In Fig.~\ref{FIG.3.}(b), the AMPT model is consistent with the experimental measurements of $S\sigma$ of proton and net proton, but slightly overestimates that of antiproton. It is believed that the nonmonotonic energy dependence of $\kappa \sigma^{2}$ indicates that critical fluctuations may occur as the systems passes through the region near the CEP~\cite{Luo:2017faz,Bazavov:2017dus}. However, the AMPT results for $\kappa \sigma^{2}$ of proton and net proton are consistent with the Poisson baseline within the large errors, as shown in Fig.~\ref{FIG.3.}(c). This is not surprising because the AMPT model does not include any critical fluctuations at the CEP. 
\begin{figure*}[htbp]
\centering
\includegraphics
[width=17.5cm]{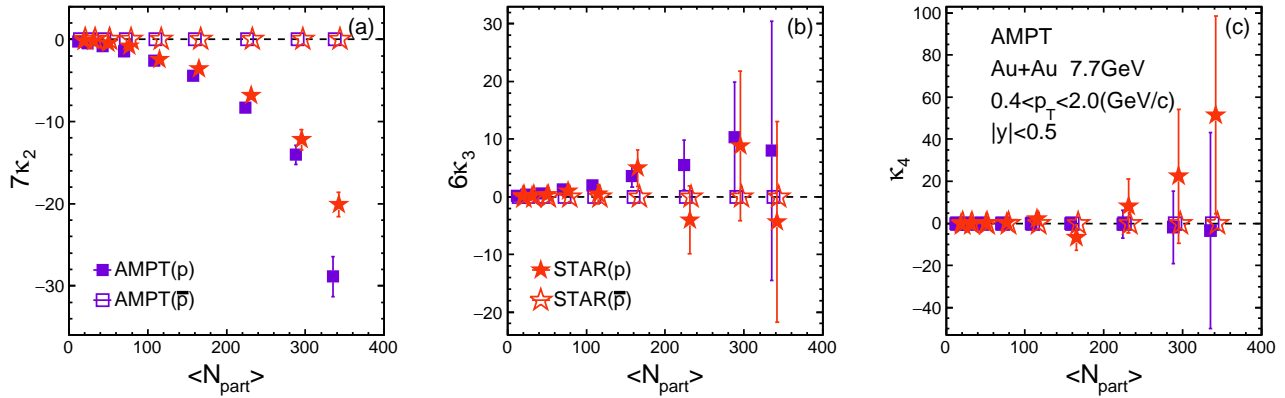}
\caption{(Color online) The AMPT results on correlation functions of proton and antiproton as a function of $\left \langle N_{part} \right \rangle$ in Au+Au collisions at $\sqrt{s_{NN}} = 7.7$ GeV, in comparisons with the STAR measurements~\cite{STAR:2021iop}.}
\label{FIG.4.}
\end{figure*}
\begin{figure*}[htbp]
\centering
\includegraphics
[width=17.5cm]{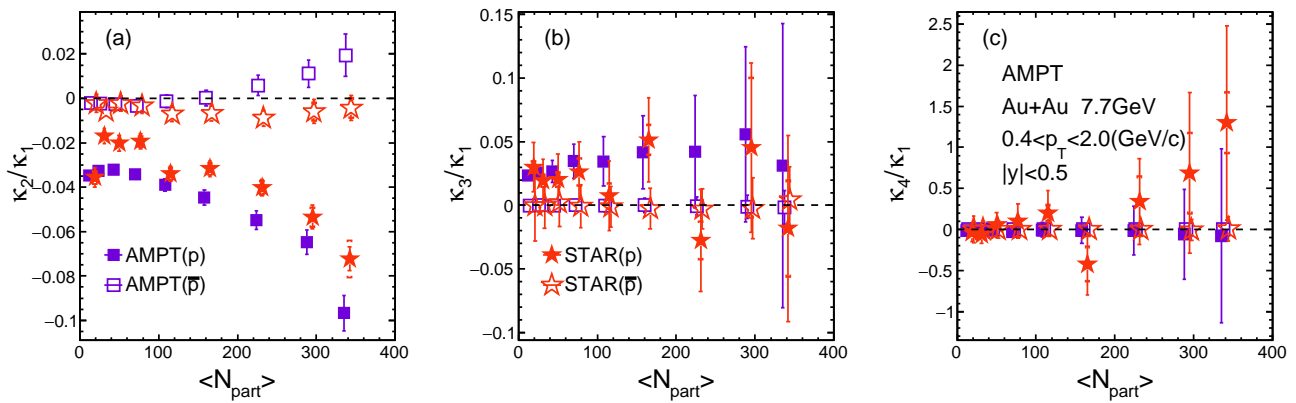}
\caption{(Color online) The AMPT results on normalized correlation functions of proton and antiproton as a function of $\left \langle N_{part} \right \rangle$ in Au+Au collisions at $\sqrt{s_{NN}} = 7.7$ GeV, in comparisons with the STAR measurements~\cite{STAR:2021iop}.}
\label{FIG.5.}
\end{figure*}

Figure~\ref{FIG.4.} shows the $\left \langle N_{part} \right \rangle$ dependence of correlation functions of proton and antiproton multiplicity distributions in Au+Au collisions at $\sqrt{s_{NN}} = 7.7$ GeV, which can be obtained according to  Eq.~(\ref{MDIV0}). The STAR data points are also deduced from the measured $C_{n}$ according to Eq.~(\ref{MDIV0}). Note that we do not show the correlation functions for net protons, because it is a bit more complicated due to the fact that net protons are composed of two kinds of particles, protons and antiprotons~\cite{Bzdak:2019pkr,Lin:2017xkd}. Because the yield of antiprotons is much lower than that of protons, the fluctuations of protons should be similar to those of net protons in Au+Au collisions at $\sqrt{s_{NN}} = 7.7$ GeV. This has been demonstrated by the fact that cumulants and cumulant ratios of proton and net protons basically overlap in Figs.~\ref{FIG.2.} and \ref{FIG.3.}. In Fig.~\ref{FIG.4.}(a), consistent with the measured two-proton correlation, the AMPT model shows a significant negative two-proton correlation, and the strength increases with the increase of $\left \langle N_{part} \right \rangle$.  In Fig.~\ref{FIG.4.}(b), the AMPT model shows a positive three-proton correlation which increases with $\left \langle N_{part} \right \rangle$, which looks consistent with the experimental measurement within large uncertainties. In Fig.~\ref{FIG.4.}(c), the four-proton correlation from the AMPT events is very small, consistent with zero. These $\left \langle N_{part} \right \rangle$ dependences of multiproton correlation functions are actually the result of the evolution of baryon stopping under the baryon number conservation, which will be discussed in Sec.~\ref{sec:partC}.

As shown in Eqs.~(\ref{MDIV5}), the cumulant ratios are actually related to normalized correlation functions. Figure~\ref{FIG.5.} shows the $\left \langle N_{part} \right \rangle$ dependence of normalized correlation functions of proton and antiproton multiplicity distributions in Au+Au collisions at $\sqrt{s_{NN}} = 7.7$ GeV. In Fig.~\ref{FIG.5.}(a), as the $\left \langle N_{part} \right \rangle$ increases, the normalized two-proton correlation decreases with a stronger strength than the experimental data, which is consistent with what has been observed in Fig.~\ref{FIG.4.}(a). However, the AMPT model shows a slightly increasing positive normalized two-antiproton correlation, which seems to be somewhat inconsistent with the experimental measurement showing a nearly zero or slightly negative correlation. We have checked that the positive two-antiproton correlation is caused by the simple coalescence scheme in our model, which disappears in the AMPT model with a new coalescence scheme~\cite{He:2017tla}. In Fig.~\ref{FIG.5.}(b), the AMPT result on $\kappa _{3}/\kappa _{1}$ for protons is positive and increases with $\left \langle N_{part} \right \rangle$, but that for antiprotons is close to zero. In Fig.~\ref{FIG.5.}(c), the AMPT results on $\kappa _{4}/\kappa _{1}$ for protons and antiprotons are both consistent with zero. Due to the large uncertainties of experimental data points, we can not conclude if there is any inconsistency between the AMPT model and the experimental measurements.

\subsection{Rapidity dependence}
\label{sec:partC}
\begin{figure}[htb]
\centering
\includegraphics
[width=9cm]{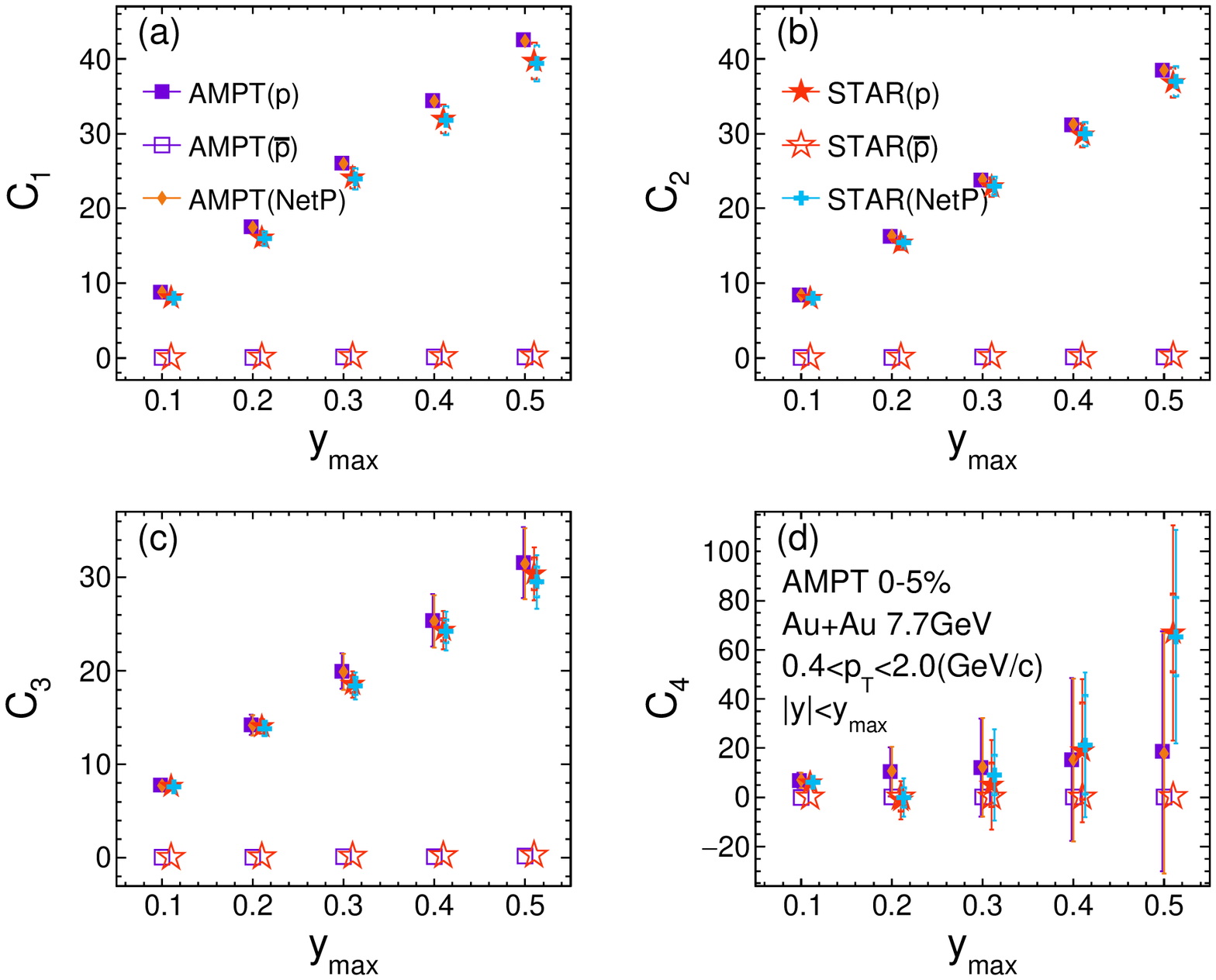}
\caption{(Color online) The AMPT results on cumulants $C_{n}$ of proton, antiproton, and net-proton distributions as a function of rapidity cut $y_{\rm max}$ in 0--5 \% central Au+Au collisions at $\sqrt{s_{NN}} = 7.7$ GeV, in comparisons with the STAR measurements~\cite{STAR:2021iop}.}
\label{FIG.6.}
\end{figure}
\begin{figure*}[htbp]
\centering
\includegraphics
[width=17.5cm]{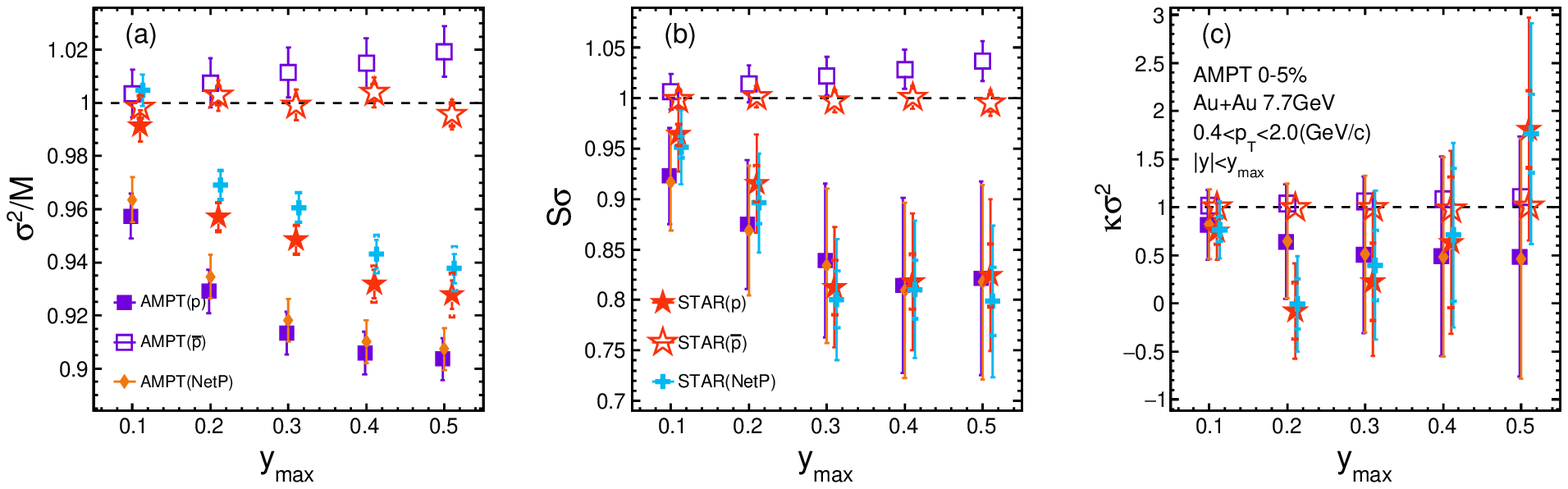}
\caption{(Color online) The AMPT results on cumulant ratios of proton, antiproton, and net-proton distributions as a function of rapidity cut $y_{\rm max}$ in 0--5 \% central Au+Au collisions at $\sqrt{s_{NN}} = 7.7$ GeV, in comparisons with the STAR measurements~\cite{STAR:2021iop}.}
\label{FIG.7.}
\end{figure*}    
\begin{figure*}[htbp]
\centering
\includegraphics
[width=17.5cm]{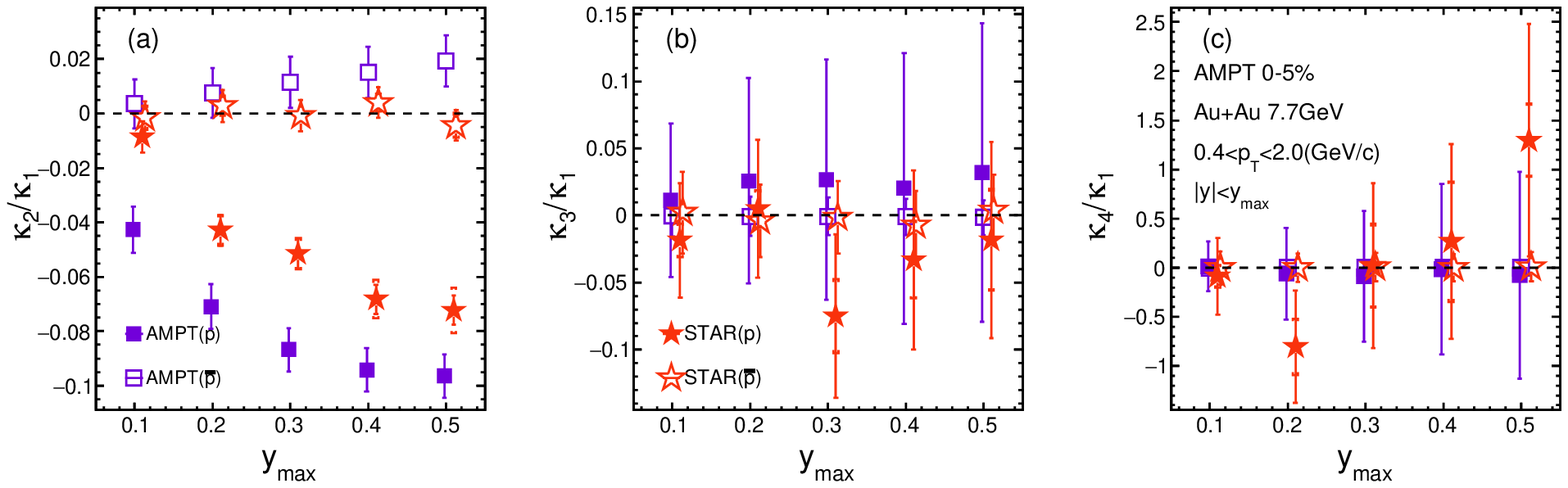}
\caption{(Color online) The AMPT results on normalized correlation functions of proton and antiproton distributions as a function of rapidity cut $y_{\rm max}$ in 0--5 \% central Au+Au collisions at $\sqrt{s_{NN}} = 7.7$ GeV, in comparisons with the STAR measurements~\cite{STAR:2021iop}.}
\label{FIG.8.}
\end{figure*}
Figure~\ref{FIG.6.} shows the rapidity cut $y_{\rm max}$ dependence of cumulants $C_{n}$ ($n=1$, 2, 3, and 4) of proton, antiproton, and net-proton multiplicity distributions in 0--5 \% central Au+Au collisions at $\sqrt{s_{NN}} = 7.7$ GeV. The rapidity of particle $y$ is required to be within $\left | y \right |<y_{\rm max}$, and the rapidity acceptance window is $\Delta y=2y_{\rm max}$. Therefore, increasing $y_{\rm max}$ actually expands the size of the rapidity acceptance window. In this analysis, the transverse momentum of particle is between 0.4 and 2.0 GeV/$c$. We find that the $C_{n}$ of protons and net protons are approximately equal due to the very small yield of antiprotons. The cumulants $C_{n}$ ($n=1$, 2, 3, and 4) of proton, antiproton, and net-proton increase with the rapidity acceptance. We observe that the AMPT results are generally in agreement with the experimental data within the errors.
       
Figure~\ref{FIG.7.} shows the rapidity cut $y_{\rm max}$ dependence of cumulant ratios of proton, antiproton, and net-proton multiplicity distributions [$C_{2}/C_{1}\left ( \sigma ^{2}/M \right )$, $C_{3}/C_{2}\left ( S\sigma \right )$ and $C_{4}/C_{2}\left ( \kappa \sigma^{2} \right )$] in 0--5 \% central Au+Au collisions at $\sqrt{s_{NN}} = 7.7$ GeV. In Fig.\ref{FIG.7.}(a) and (b), the AMPT results show consistent trends with experimental measurements, but the AMPT results of $\sigma ^{2}/M$ for protons and net protons are slightly smaller than the experimental data. In Fig.\ref{FIG.7.}(c), the AMPT results show a decreasing trend for protons and net protons. According to Eqs.~(\ref{MDIV5}), we know that the cumulant ratios are related to the interplay between different orders of normalized correlation functions $\kappa _{n}/\kappa _{1}$. We observe that the cumulant ratios $\sigma ^{2}/M$ , $S\sigma$, and $\kappa \sigma^{2}$ of protons deviate from the Poisson baseline, which will be shown later to be a result of negative two-proton correlation. 

Figure~\ref{FIG.8.} shows the rapidity cut $y_{\rm max}$ dependence of normalized correlation functions of proton and antiproton distributions in 0--5 \% central Au+Au collisions at $\sqrt{s_{NN}} = 7.7$ GeV. In Fig.~\ref{FIG.8.}(a), the $\kappa _{2}/\kappa _{1}$ for protons is negative, which decreases monotonically with the increase of rapidity acceptance. However, we observe that the AMPT results on $\kappa _{2}/\kappa _{1}$ for protons have a stronger strength than the experimental data. In Fig.~\ref{FIG.8.}(b) and (c), the AMPT results on $\kappa _{3}/\kappa _{1}$ and $\kappa _{4}/\kappa _{1}$ for protons and antiprotons are both close to zero, which can not produce a rapidity-dependent $\kappa _{4}/\kappa _{1}$ or long-range four-proton correlation~\cite{Bzdak:2016sxg,Bzdak:2017ltv,Ling:2015yau}.  

\subsection{Transverse momentum dependence}
\label{sec:partC}
\begin{figure}[htb]
\centering
\includegraphics
[width=9cm]{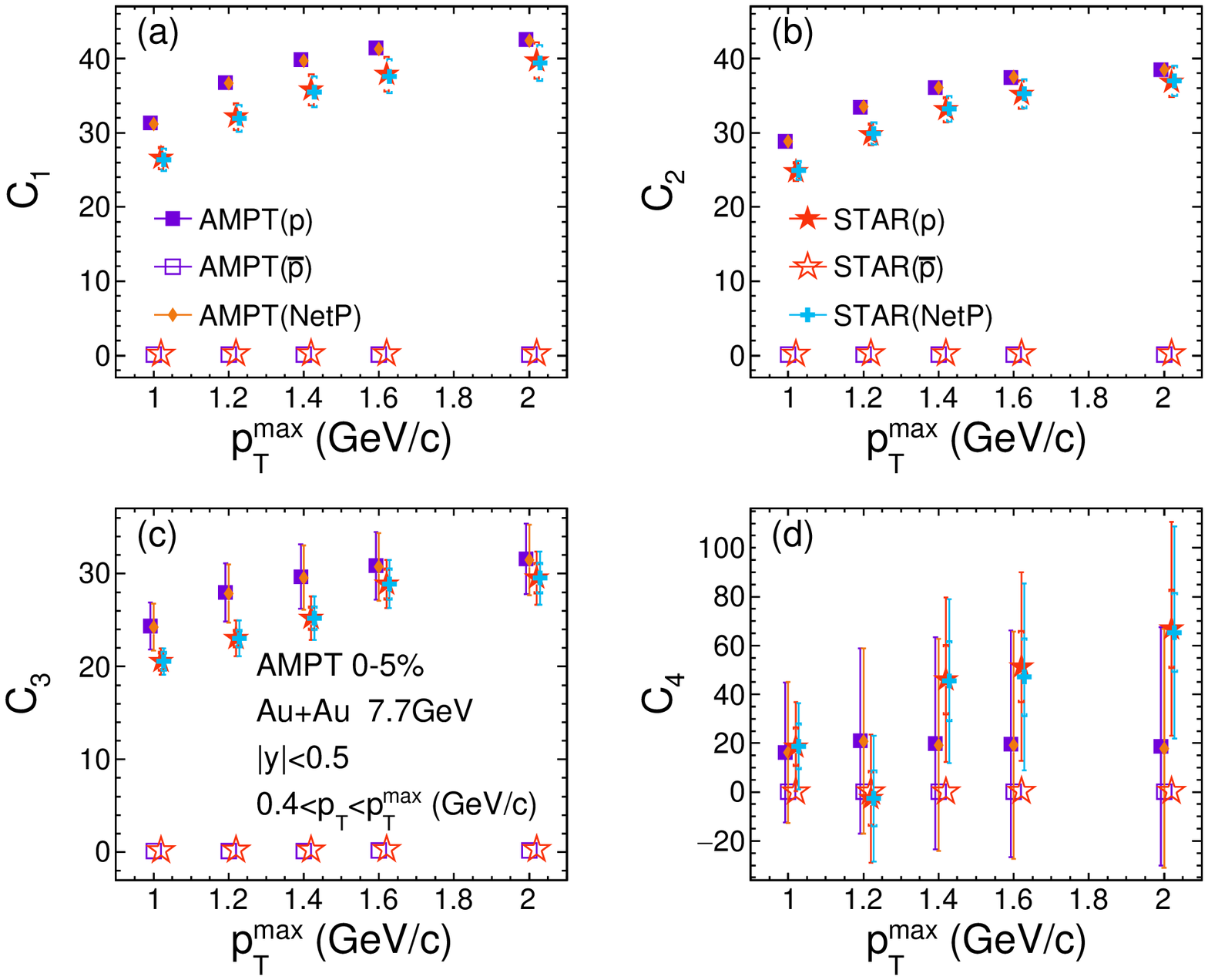}
\caption{(Color online) The AMPT results on cumulants $C_{n}$ of proton, antiproton, and net-proton distributions as a function of transverse momentum cut $p_{T}^{\rm max}$ in 0--5 \% central Au+Au collisions at $\sqrt{s_{NN}} = 7.7$ GeV, in comparisons with the STAR measurements~\cite{STAR:2021iop}.}
\label{FIG.9.}
\end{figure}
\begin{figure*}[htbp]
\centering
\includegraphics
[width=17.5cm]{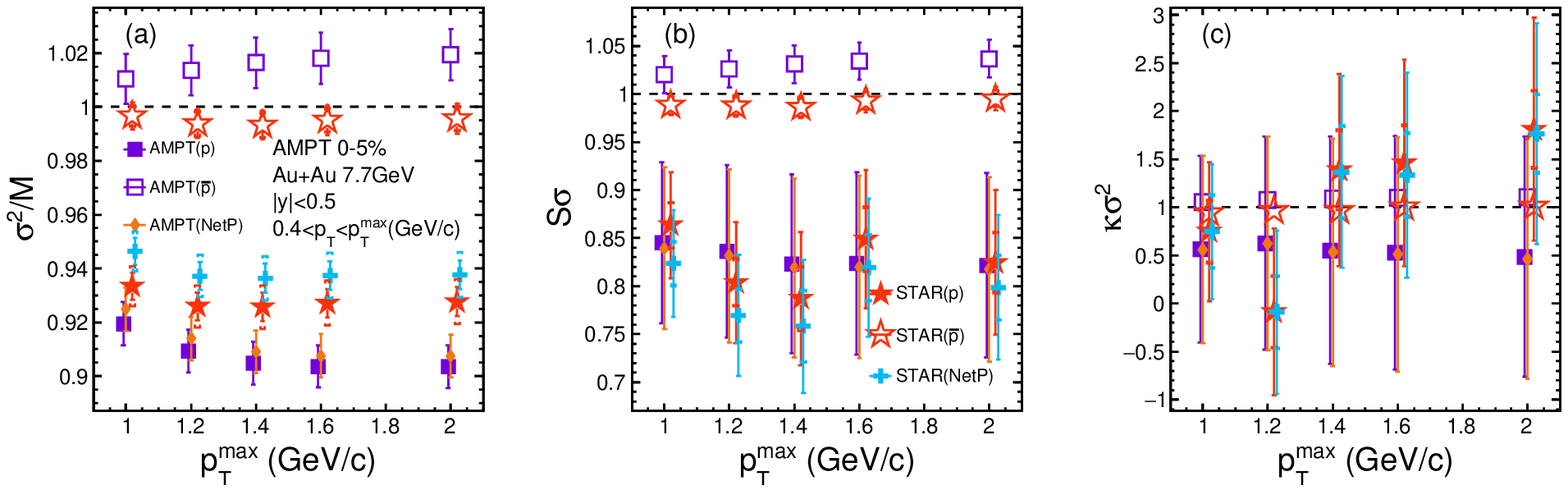}
\caption{(Color online) The AMPT results on cumulant ratios of proton, antiproton, and net-proton distributions as a function of transverse momentum cut $p_{T}^{\rm max}$ in 0--5 \% central Au+Au collisions at $\sqrt{s_{NN}} = 7.7$ GeV, in comparisons with the STAR measurements~\cite{STAR:2021iop}.}
\label{FIG.10.}
\end{figure*}      
\begin{figure*}[htbp]
\centering
\includegraphics
[width=17.5cm]{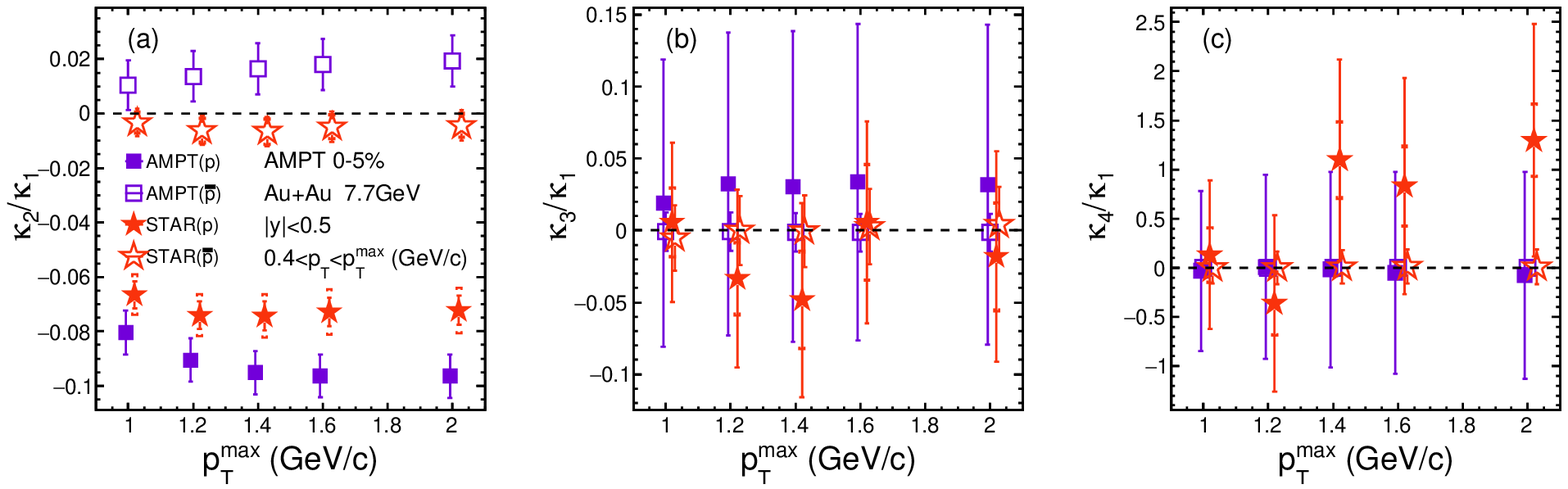}
\caption{(Color online) The AMPT results on normalized correlation functions of proton and antiproton distributions as a function of transverse momentum cut $p_{T}^{\rm max}$ in 0--5 \% central Au+Au collisions at $\sqrt{s_{NN}} = 7.7$ GeV, in comparisons with the STAR measurements~\cite{STAR:2021iop}.}
\label{FIG.11.}
\end{figure*}
Figure~\ref{FIG.9.} shows transverse momentum cut $p_{T}^{\rm max}$ dependence of cumulants $C_{n}$ ($n=1$, 2, 3, and 4) of proton, antiproton, and net-proton multiplicity distributions in 0--5 \% central Au+Au collisions at $\sqrt{s_{NN}} = 7.7$ GeV. The analysis are performed for the particles in the transverse momentum range of the $0.4<p_{T}<p_{T}^{\rm max}$ GeV/$c$ within the rapidity window of $\left | y \right |<0.5$. As the upper transverse momentum cut increases from 1.0 GeV/$c$ to 2.0 GeV/$c$, the AMPT results on the cumulants $C_{n}$ ($n=1$, 2, and 3) of proton, antiproton, and net-proton increases, which are consistent with the experimental data. However, we observe that the AMPT results of $C_{4}$ always stay around 20 without any $p_{T}^{\rm max}$ dependence. 

Figure~\ref{FIG.10.} shows the transverse momentum cut $p_{T}^{\rm max}$ dependence of cumulant ratios of proton, antiproton, and net-proton multiplicity distributions [$C_{2}/C_{1}\left ( \sigma ^{2}/M \right )$, $C_{3}/C_{2}\left ( S\sigma \right )$, and $C_{4}/C_{2}\left ( \kappa \sigma^{2} \right )$] in 0--5 \% central Au+Au collisions at $\sqrt{s_{NN}} = 7.7$ GeV. In Fig.~\ref{FIG.10.}(a) and (b), the AMPT results on both $\sigma ^{2}/M$ and $S\sigma$ show a weak dependence on the $p_{T}$ acceptance for proton, antiproton, and net-proton, which is  consistent the trend from the STAR data. In Fig.\ref{FIG.10.}(c), the AMPT results on the cumulant ratio of $\kappa \sigma^{2}$ are systematically below unity for  both protons and net protons within large uncertainties.
       	
Figure~\ref{FIG.11.} shows the transverse momentum cut $p_{T}^{\rm max}$ dependence of normalized correlation functions of proton and antiproton distributions in 0--5 \% central Au+Au collisions at $\sqrt{s_{NN}} = 7.7$ GeV. In Fig.~\ref{FIG.11.}(a), the AMPT result on the $\kappa _{2}/\kappa _{1}$ for protons decreases with increasing the $p_{T}$ acceptance and is slightly more negative than the experimental data. In Fig.~\ref{FIG.11.}(b) and (c), the AMPT results on both $\kappa _{3}/\kappa _{1}$ and $\kappa _{4}/\kappa _{1}$ are consistent with zero. Our results about both rapidity and transverse momentum acceptance effects are consistent with the recent UrQMD calculations as a result of baryon number conservation~\cite{He:2017zpg}. 

\subsection{Stage evolution}
\label{sec:partC}
Relativistic heavy-ion collisions are actually a complex dynamical evolution including several important evolution stages. In order to understand the dynamics of the fluctuation observables, it is necessary to compare cumulants and correlation functions at each evolution stage. As introduced in Sec.~\ref{ampt},  the AMPT model is a hybrid model consisting of four components, corresponding to four evolution stages. Taking advantage of the AMPT model, we focus on the dynamical development of cumulants and correlation functions for four evolution stages in Au+Au collisions at $\sqrt{s_{NN}} = 7.7$ GeV, i.e., (1) ``Initial state''  refers to the initially formed partonic matter consisting quarks and antiquarks, (2) ``After parton cascade'' refers to the finial freeze-out partonic matter consisting quarks and antiquarks that have undergone parton cascade, (3) ``After hadronization'' refers to the newly formed hadronic matter transformed from the freeze-out partonic matter through the hadronization of coalescence,  and (4) ``Final state''  refers to the final freeze-out hadronic matter which has undergone hadronic rescatterings and resonance decays.
\begin{figure*}[htbp]
\centering
\includegraphics
[width=14.5cm]{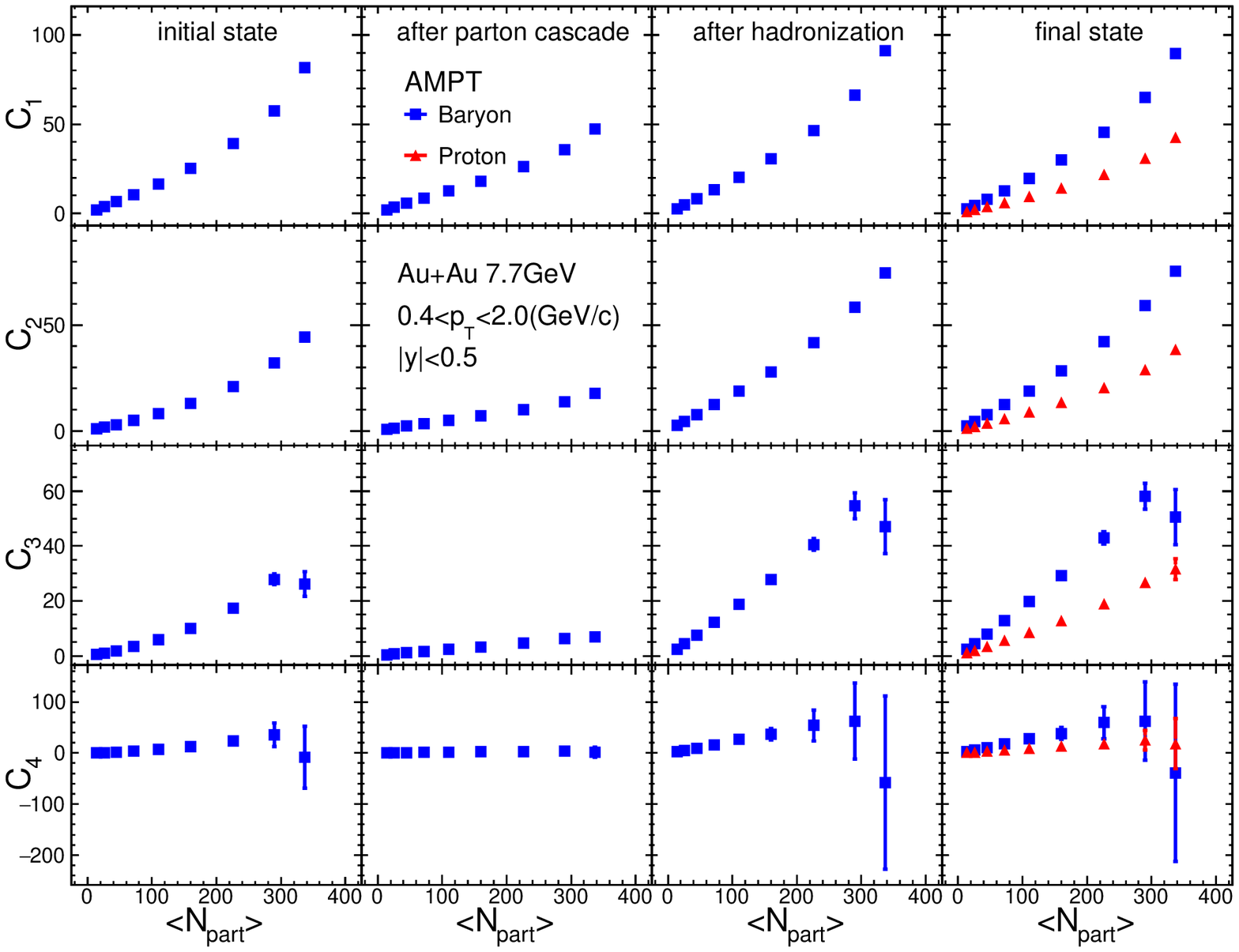}
\caption{The AMPT results on cumulants $C_{n}$ of baryon distributions as a function of $\left \langle N_{part} \right \rangle$ at different evolution stages in Au+Au collisions at $\sqrt{s_{NN}} = 7.7$ GeV.}
\label{FIG.12.}
\end{figure*}

To see the stage evolution of cumulants of protons, Fig.~\ref{FIG.12.} shows the $\left \langle N_{part} \right \rangle$ dependence of cumulants of baryon multiplicity distributions at four different evolution stages (from left to right) in Au+Au collisions at $\sqrt{s_{NN}} = 7.7$ GeV. Note that because the baryon number is carried by quarks, we consider the quarks in the first two stages as one-third of baryons. We observe that all orders of cumulants always increase with $\left \langle N_{part} \right \rangle$. The cumulants decrease after parton cascade, compared to those in the initial state. The hadronization process  increases the cumulants of baryons, relative to those after parton cascade. However, hadronic rescatterings have little effect on the cumulants of baryons. Let us discuss the stage evolution of the average number of baryons $C_{1}$ as an example. The number of baryons decreases after parton cascade, because the baryon distribution is diffused by parton cascade. Hadronization increases the number of baryons, because coalescence brings baryon number into the acceptance window. However, the role of baryon diffusion in the hadronic phase appears weak. At the same time, we find that the centrality dependence of cumulants of protons is similar to that of baryons but with smaller magnitudes in the final state. Note that we apply a same $p_T$ cut ($0.4<p_T<2$ GeV/$c$) for all stages.  However, we test a different cut ($0.4/3<p_T<2/3$ GeV/$c$) for quarks for the first two stages, if we simply assume that a baryon with $p_T$ is made up of three quarks with $p_T/3$. As it turns out, the evolution trends are still there.

\begin{figure*}[htbp]
\centering
\includegraphics
[width=14.5cm]{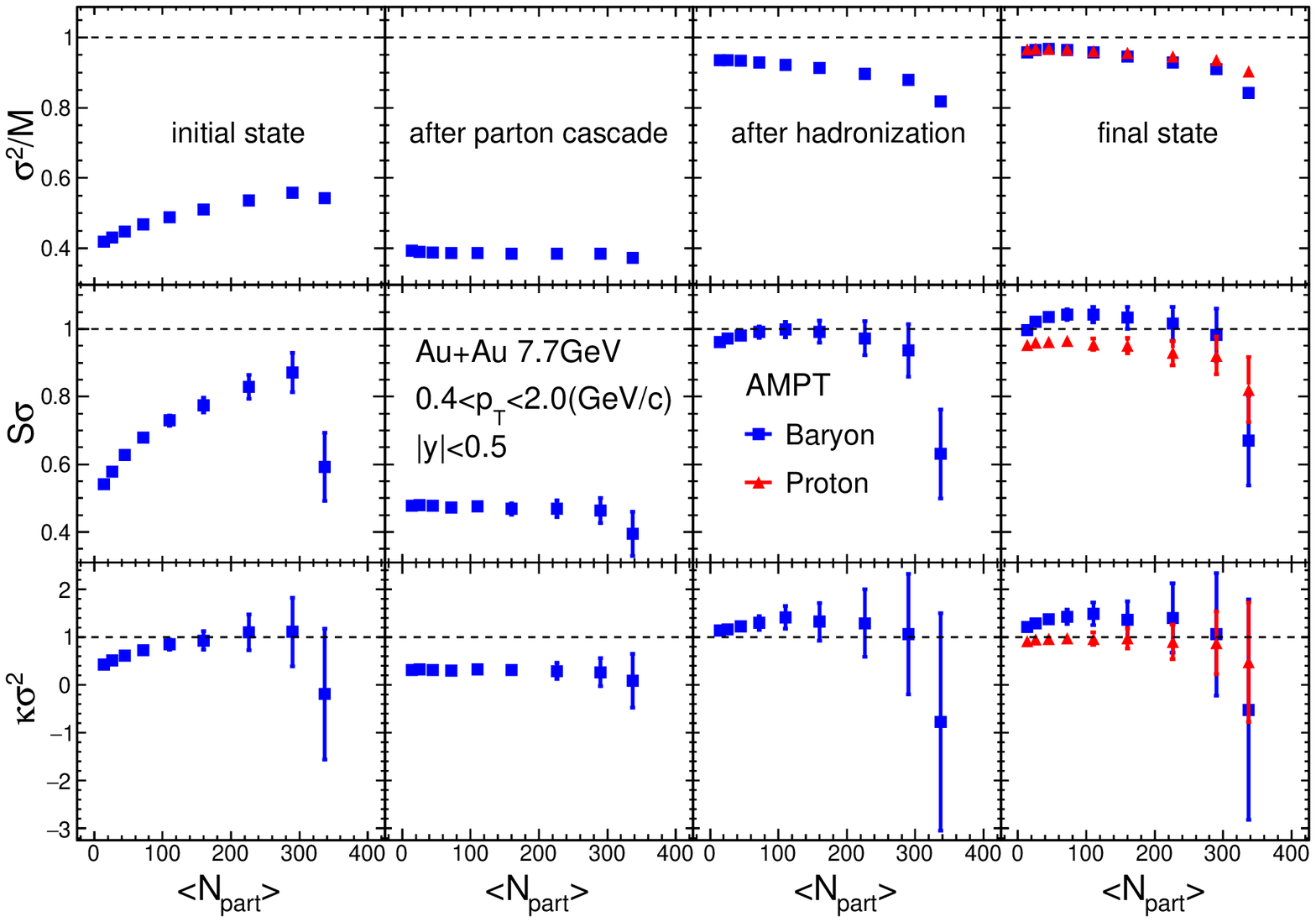}
\caption{The AMPT results on cumulant ratios of baryon distributions as a function of $\left \langle N_{part} \right \rangle$ at different evolution stages in Au+Au collisions at $\sqrt{s_{NN}} = 7.7$ GeV.}
\label{FIG.13.}
\end{figure*}

Figure~\ref{FIG.13.} shows the $\left \langle N_{part} \right \rangle$ dependence of cumulant ratios of baryon multiplicity distributions at four different evolution stages in Au+Au collisions at $\sqrt{s_{NN}} = 7.7$ GeV. As shown in Eqs.~(\ref{MDIV5}), all Poisson baselines for the three cumulant ratios will be unity, if there are not any multiparticle correlations. The AMPT results on $\sigma ^{2}/M$ of baryons are always lower than the Poisson baseline for four different evolution stages, indicating that there is always a negative two-baryon correlation throughout the evolution of heavy-ion collisions. For $S\sigma$ and $\kappa \sigma^{2}$, they are more complicated than $\sigma ^{2}/M$, because they consist of both two-baryon correlation and multibaryon correlations. The clean information about different multibaryon correlations will be discussed next. However, it is certain that the cumulant ratios develop with the evolution of heavy-ion collisions. Therefore, it is important to include the dynamical evolution effect on cumulant ratios to search for the possible critical fluctuations at the CEP. In addition, we observe that the cumulant ratios of protons are similar to those of baryons in trend, with slightly different magnitudes in the final state.

\begin{figure*}[htbp]
\centering
\includegraphics
[width=14.5cm]{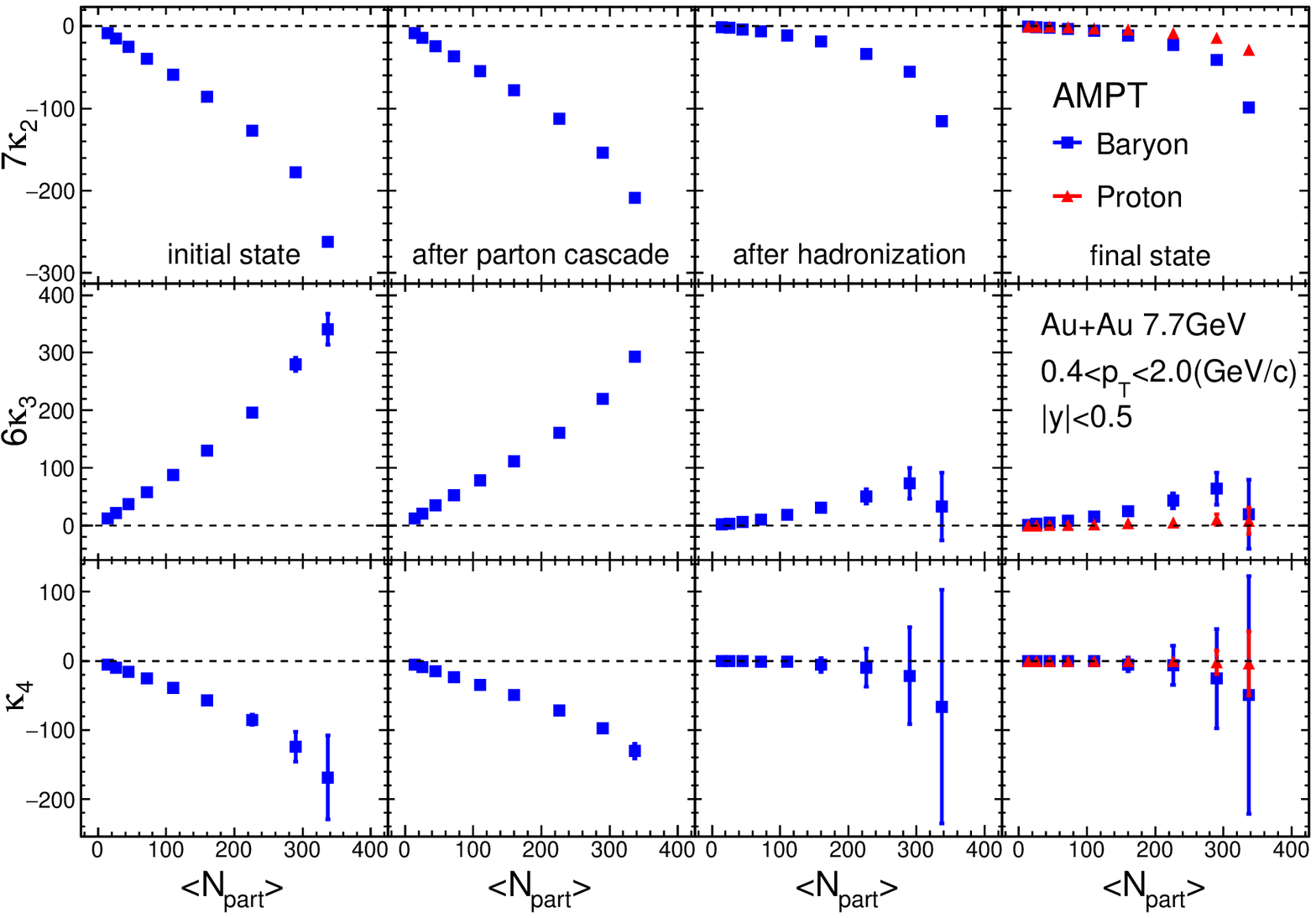}
\caption{The AMPT results on correlation functions of baryons as a function of $\left \langle N_{part} \right \rangle$ at different evolution stages in Au+Au collisions at $\sqrt{s_{NN}} = 7.7$ GeV.}
\label{FIG.14.}
\end{figure*}

Multiparticle correlation functions are much cleaner than cumulants and cumulant ratios~\cite{Bzdak:2019pkr}. Figure~\ref{FIG.14.} shows the $\left \langle N_{part} \right \rangle$ dependence of $n$-baryon correlation $\kappa_{n}$ at four different evolution stages in Au+Au collisions at $\sqrt{s_{NN}} = 7.7$ GeV. We observe negative two-baryon correlations, positive three-baryon correlations, and negative four-baryon correlations in the initial state. The signs are consistent with the expectation of baryon number conservation. For multibaryon correlations, the correlation strength increases with $\left \langle N_{part} \right \rangle$, which indicates that more baryons are stopped into the midrapidity region in more central collisions due to baryon stopping. Furthermore, we observe that the multibaryon correlations are getting weaker with the stage evolution of heavy-ion collisions. It indicates that multibaryon correlations are weakened, but has persisted throughout the evolution of heavy-ion collisions. We observe that multiproton correlations are similar to multibaryon correlations in trend, but with smaller strengths, suggesting that protons can be considered to some extent as a proxy of baryons for measuring multibaryon correlations.

\begin{figure}[htb]
\centering
\includegraphics
[width=8.5cm]{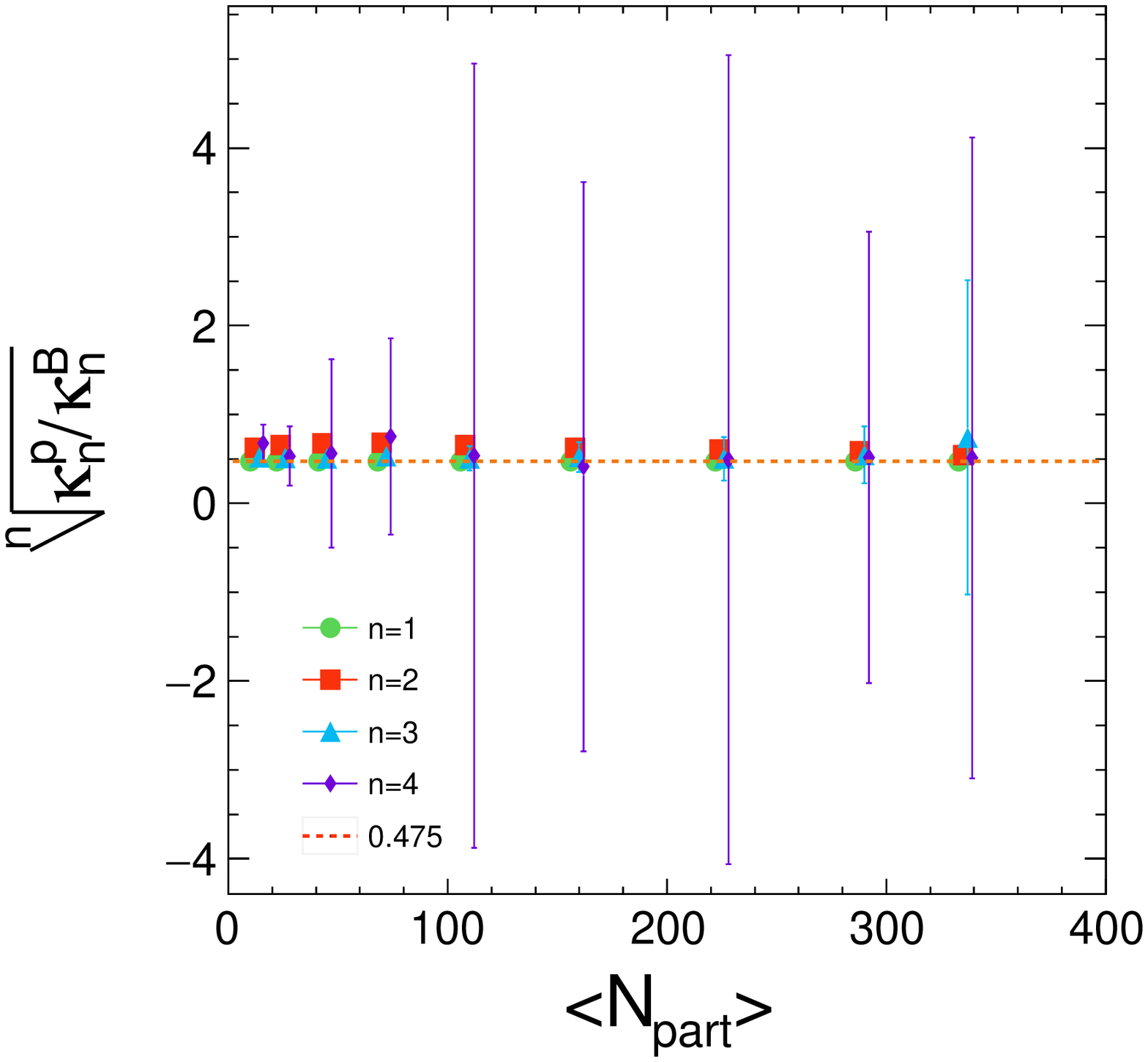}
\caption{(Color online) The AMPT results on the acceptance factor $\sqrt[n]{\kappa^{p}_{n}/\kappa^{B}_{n}}$ as a function of $\left \langle N_{part} \right \rangle$ in Au+Au collisions at $\sqrt{s_{NN}} = 7.7$ GeV.}
\label{FIG.15.}
\end{figure}

According to Refs.~\cite{Vovchenko:2021kxx,Kitazawa:2011wh,Kitazawa:2012at,Bzdak:2012ab}, multibaryon correlation and multi-proton correlation are related by $\kappa^{B}_{n}$= $\kappa^{p}_{n}$/$q^{n}$, where $q$ is an effective acceptance factor representing the proton fraction of baryons within limited acceptance and efficiency. This relation is also consistent with the expectation from baryon number conservation if $\left \langle N^{B} \right \rangle$= $\left \langle N^{p} \right \rangle$/$q$, as shown in Eq.~(\ref{corrbnc}).  We extract the acceptance factors $q$ by calculating the $n$th roots of the ratios of $n$-proton correlation to $n$-baryon correlation, i.e., $\sqrt[n]{\kappa^{p}_{n}/\kappa^{B}_{n}}$. Figure~\ref{FIG.15.} shows the acceptance factor $\sqrt[n]{\kappa^{p}_{n}/\kappa^{B}_{n}}$ as a function of $\left \langle N_{part} \right \rangle$ in Au+Au collisions at $\sqrt{s_{NN}} = 7.7$ GeV. We find that the acceptance factor is almost independent of centrality and is about 0.475 by a constant fitting, which is slightly different from 1/2~\cite{Kitazawa:2011wh,Kitazawa:2012at}. From Figs. ~\ref{FIG.14.} and ~\ref{FIG.15.}, it can be seen that most of the multiproton correlations or multibaryon correlations should be due to the same source, i.e., baryon number conservation, in the AMPT model. The acceptance factor can reflect the degree to which protons can act as a proxy for baryons in baryon number conservation. However, the relation between baryon and proton number fluctuations may be more complicated if there are many other contributing sources, especially critical fluctuations, which need to be investigated furthermore.

\section{Summary}
\label{framework}

In summary, the centrality, rapidity and transverse momentum dependences of cumulants and correlation functions of proton, antiproton, and net-proton multiplicity distributions have been studied in Au+Au Collisions at $\sqrt{s_{NN}} = 7.7$ GeV using a multiphase transport model. Our results of cumulants, cumulant ratios, correlation functions and normalized correlation functions of proton, antiproton, and net-proton multiplicity distributions basically describe the trend in the experimental data. The AMPT results are consistent with the expectation from baryon number conservation. By focusing on the dynamical stage evolution of cumulants and multiparticle correlation functions, we find that multibaryon (proton) correlations due to baryon number conservation are diluted with the evolution of heavy-ion collisions. Since the physics of the QCD critical fluctuations is not included in the AMPT model, our results are expected to only provide a baseline for searching for the possible critical behaviors at the CEP in relativistic heavy-ion collisions.

\section*{ACKNOWLEDGMENTS}
We thank Prof. Xiao-Feng Luo for helpful discussions, and Dr. Chen Zhong for maintaining the high-quality performance of the Fudan supercomputing platform for nuclear physics. This work is supported by the National Natural Science Foundation of China under Contracts No.12147101, No. 11961131011, No. 11890710, No. 11890714, No. 11835002, the Strategic Priority Research Program of Chinese Academy of Sciences under Grant No. XDB34030000, and the Guangdong Major Project of Basic and Applied Basic Research under Grant No. 2020B0301030008.

\bibliography{myref}

\begin{thebibliography}{85}
\expandafter\ifx\csname natexlab\endcsname\relax\def\natexlab#1{#1}\fi
\expandafter\ifx\csname bibnamefont\endcsname\relax
  \def\bibnamefont#1{#1}\fi
\expandafter\ifx\csname bibfnamefont\endcsname\relax
  \def\bibfnamefont#1{#1}\fi
\expandafter\ifx\csname citenamefont\endcsname\relax
  \def\citenamefont#1{#1}\fi
\expandafter\ifx\csname url\endcsname\relax
  \def\url#1{\texttt{#1}}\fi
\expandafter\ifx\csname urlprefix\endcsname\relax\def\urlprefix{URL }\fi
\providecommand{\bibinfo}[2]{#2}
\providecommand{\eprint}[2][]{\url{#2}}

\bibitem[{\citenamefont{Ding et~al.}(2015)\citenamefont{Ding, Karsch, and
  Mukherjee}}]{Ding:2015ona}
\bibinfo{author}{\bibfnamefont{H.-T.} \bibnamefont{Ding}},
  \bibinfo{author}{\bibfnamefont{F.}~\bibnamefont{Karsch}}, \bibnamefont{and}
  \bibinfo{author}{\bibfnamefont{S.}~\bibnamefont{Mukherjee}},
  \bibinfo{journal}{Int. J. Mod. Phys. E} \textbf{\bibinfo{volume}{24}},
  \bibinfo{pages}{1530007} (\bibinfo{year}{2015}), \eprint{1504.05274}.

\bibitem[{\citenamefont{Bazavov et~al.}(2014)}]{HotQCD:2014kol}
\bibinfo{author}{\bibfnamefont{A.}~\bibnamefont{Bazavov}} \bibnamefont{et~al.}
  (\bibinfo{collaboration}{HotQCD}), \bibinfo{journal}{Phys. Rev. D}
  \textbf{\bibinfo{volume}{90}}, \bibinfo{pages}{094503}
  (\bibinfo{year}{2014}), \eprint{1407.6387}.

\bibitem[{\citenamefont{Bellwied et~al.}(2013)\citenamefont{Bellwied, Borsanyi,
  Fodor, Katz, and Ratti}}]{Bellwied:2013cta}
\bibinfo{author}{\bibfnamefont{R.}~\bibnamefont{Bellwied}},
  \bibinfo{author}{\bibfnamefont{S.}~\bibnamefont{Borsanyi}},
  \bibinfo{author}{\bibfnamefont{Z.}~\bibnamefont{Fodor}},
  \bibinfo{author}{\bibfnamefont{S.~D.} \bibnamefont{Katz}}, \bibnamefont{and}
  \bibinfo{author}{\bibfnamefont{C.}~\bibnamefont{Ratti}},
  \bibinfo{journal}{Phys. Rev. Lett.} \textbf{\bibinfo{volume}{111}},
  \bibinfo{pages}{202302} (\bibinfo{year}{2013}), \eprint{1305.6297}.

\bibitem[{\citenamefont{Borsanyi et~al.}(2014)\citenamefont{Borsanyi, Fodor,
  Katz, Krieg, Ratti, and Szabo}}]{Borsanyi:2014ewa}
\bibinfo{author}{\bibfnamefont{S.}~\bibnamefont{Borsanyi}},
  \bibinfo{author}{\bibfnamefont{Z.}~\bibnamefont{Fodor}},
  \bibinfo{author}{\bibfnamefont{S.~D.} \bibnamefont{Katz}},
  \bibinfo{author}{\bibfnamefont{S.}~\bibnamefont{Krieg}},
  \bibinfo{author}{\bibfnamefont{C.}~\bibnamefont{Ratti}}, \bibnamefont{and}
  \bibinfo{author}{\bibfnamefont{K.~K.} \bibnamefont{Szabo}},
  \bibinfo{journal}{Phys. Rev. Lett.} \textbf{\bibinfo{volume}{113}},
  \bibinfo{pages}{052301} (\bibinfo{year}{2014}), \eprint{1403.4576}.

\bibitem[{\citenamefont{Zhang et~al.}(2017)\citenamefont{Zhang, Hou, Kojo, and
  Qin}}]{Zhang:2017icm}
\bibinfo{author}{\bibfnamefont{H.}~\bibnamefont{Zhang}},
  \bibinfo{author}{\bibfnamefont{D.}~\bibnamefont{Hou}},
  \bibinfo{author}{\bibfnamefont{T.}~\bibnamefont{Kojo}}, \bibnamefont{and}
  \bibinfo{author}{\bibfnamefont{B.}~\bibnamefont{Qin}},
  \bibinfo{journal}{Phys. Rev. D} \textbf{\bibinfo{volume}{96}},
  \bibinfo{pages}{114029} (\bibinfo{year}{2017}), \eprint{1709.05654}.

\bibitem[{\citenamefont{Mitter et~al.}(2015)\citenamefont{Mitter, Pawlowski,
  and Strodthoff}}]{Mitter:2014wpa}
\bibinfo{author}{\bibfnamefont{M.}~\bibnamefont{Mitter}},
  \bibinfo{author}{\bibfnamefont{J.~M.} \bibnamefont{Pawlowski}},
  \bibnamefont{and}
  \bibinfo{author}{\bibfnamefont{N.}~\bibnamefont{Strodthoff}},
  \bibinfo{journal}{Phys. Rev. D} \textbf{\bibinfo{volume}{91}},
  \bibinfo{pages}{054035} (\bibinfo{year}{2015}), \eprint{1411.7978}.

\bibitem[{\citenamefont{Herbst et~al.}(2014)\citenamefont{Herbst, Mitter,
  Pawlowski, Schaefer, and Stiele}}]{Herbst:2013ufa}
\bibinfo{author}{\bibfnamefont{T.~K.} \bibnamefont{Herbst}},
  \bibinfo{author}{\bibfnamefont{M.}~\bibnamefont{Mitter}},
  \bibinfo{author}{\bibfnamefont{J.~M.} \bibnamefont{Pawlowski}},
  \bibinfo{author}{\bibfnamefont{B.-J.} \bibnamefont{Schaefer}},
  \bibnamefont{and} \bibinfo{author}{\bibfnamefont{R.}~\bibnamefont{Stiele}},
  \bibinfo{journal}{Phys. Lett. B} \textbf{\bibinfo{volume}{731}},
  \bibinfo{pages}{248} (\bibinfo{year}{2014}), \eprint{1308.3621}.

\bibitem[{\citenamefont{Fu et~al.}(2020)\citenamefont{Fu, Pawlowski, and
  Rennecke}}]{Fu:2019hdw}
\bibinfo{author}{\bibfnamefont{W.-j.} \bibnamefont{Fu}},
  \bibinfo{author}{\bibfnamefont{J.~M.} \bibnamefont{Pawlowski}},
  \bibnamefont{and} \bibinfo{author}{\bibfnamefont{F.}~\bibnamefont{Rennecke}},
  \bibinfo{journal}{Phys. Rev. D} \textbf{\bibinfo{volume}{101}},
  \bibinfo{pages}{054032} (\bibinfo{year}{2020}), \eprint{1909.02991}.

\bibitem[{\citenamefont{Fukushima}(2004)}]{Fukushima:2003fw}
\bibinfo{author}{\bibfnamefont{K.}~\bibnamefont{Fukushima}},
  \bibinfo{journal}{Phys. Lett. B} \textbf{\bibinfo{volume}{591}},
  \bibinfo{pages}{277} (\bibinfo{year}{2004}), \eprint{hep-ph/0310121}.

\bibitem[{\citenamefont{Skokov et~al.}(2011)\citenamefont{Skokov, Friman, and
  Redlich}}]{Skokov:2010uh}
\bibinfo{author}{\bibfnamefont{V.}~\bibnamefont{Skokov}},
  \bibinfo{author}{\bibfnamefont{B.}~\bibnamefont{Friman}}, \bibnamefont{and}
  \bibinfo{author}{\bibfnamefont{K.}~\bibnamefont{Redlich}},
  \bibinfo{journal}{Phys. Rev. C} \textbf{\bibinfo{volume}{83}},
  \bibinfo{pages}{054904} (\bibinfo{year}{2011}), \eprint{1008.4570}.

\bibitem[{\citenamefont{Pisarski and Skokov}(2016)}]{Pisarski:2016ixt}
\bibinfo{author}{\bibfnamefont{R.~D.} \bibnamefont{Pisarski}} \bibnamefont{and}
  \bibinfo{author}{\bibfnamefont{V.~V.} \bibnamefont{Skokov}},
  \bibinfo{journal}{Phys. Rev. D} \textbf{\bibinfo{volume}{94}},
  \bibinfo{pages}{034015} (\bibinfo{year}{2016}), \eprint{1604.00022}.

\bibitem[{\citenamefont{Li et~al.}(2018)\citenamefont{Li, Chen, Li, and
  Huang}}]{Li:2017ple}
\bibinfo{author}{\bibfnamefont{Z.}~\bibnamefont{Li}},
  \bibinfo{author}{\bibfnamefont{Y.}~\bibnamefont{Chen}},
  \bibinfo{author}{\bibfnamefont{D.}~\bibnamefont{Li}}, \bibnamefont{and}
  \bibinfo{author}{\bibfnamefont{M.}~\bibnamefont{Huang}},
  \bibinfo{journal}{Chin. Phys. C} \textbf{\bibinfo{volume}{42}},
  \bibinfo{pages}{013103} (\bibinfo{year}{2018}), \eprint{1706.02238}.

\bibitem[{\citenamefont{Bzdak et~al.}(2020)\citenamefont{Bzdak, Esumi, Koch,
  Liao, Stephanov, and Xu}}]{Bzdak:2019pkr}
\bibinfo{author}{\bibfnamefont{A.}~\bibnamefont{Bzdak}},
  \bibinfo{author}{\bibfnamefont{S.}~\bibnamefont{Esumi}},
  \bibinfo{author}{\bibfnamefont{V.}~\bibnamefont{Koch}},
  \bibinfo{author}{\bibfnamefont{J.}~\bibnamefont{Liao}},
  \bibinfo{author}{\bibfnamefont{M.}~\bibnamefont{Stephanov}},
  \bibnamefont{and} \bibinfo{author}{\bibfnamefont{N.}~\bibnamefont{Xu}},
  \bibinfo{journal}{Phys. Rept.} \textbf{\bibinfo{volume}{853}},
  \bibinfo{pages}{1} (\bibinfo{year}{2020}), \eprint{1906.00936}.

\bibitem[{\citenamefont{Luo et~al.}(2020)\citenamefont{Luo, Shi, Xu, and
  Zhang}}]{Luo:2020pef}
\bibinfo{author}{\bibfnamefont{X.}~\bibnamefont{Luo}},
  \bibinfo{author}{\bibfnamefont{S.}~\bibnamefont{Shi}},
  \bibinfo{author}{\bibfnamefont{N.}~\bibnamefont{Xu}}, \bibnamefont{and}
  \bibinfo{author}{\bibfnamefont{Y.}~\bibnamefont{Zhang}},
  \bibinfo{journal}{Particles} \textbf{\bibinfo{volume}{3}},
  \bibinfo{pages}{278} (\bibinfo{year}{2020}), \eprint{2004.00789}.

\bibitem[{\citenamefont{Aoki et~al.}(2006)\citenamefont{Aoki, Endrodi, Fodor,
  Katz, and Szabo}}]{Aoki:2006we}
\bibinfo{author}{\bibfnamefont{Y.}~\bibnamefont{Aoki}},
  \bibinfo{author}{\bibfnamefont{G.}~\bibnamefont{Endrodi}},
  \bibinfo{author}{\bibfnamefont{Z.}~\bibnamefont{Fodor}},
  \bibinfo{author}{\bibfnamefont{S.~D.} \bibnamefont{Katz}}, \bibnamefont{and}
  \bibinfo{author}{\bibfnamefont{K.~K.} \bibnamefont{Szabo}},
  \bibinfo{journal}{Nature} \textbf{\bibinfo{volume}{443}},
  \bibinfo{pages}{675} (\bibinfo{year}{2006}), \eprint{hep-lat/0611014}.

\bibitem[{\citenamefont{Aoki et~al.}(2009)\citenamefont{Aoki, Borsanyi, Durr,
  Fodor, Katz, Krieg, and Szabo}}]{Aoki:2009sc}
\bibinfo{author}{\bibfnamefont{Y.}~\bibnamefont{Aoki}},
  \bibinfo{author}{\bibfnamefont{S.}~\bibnamefont{Borsanyi}},
  \bibinfo{author}{\bibfnamefont{S.}~\bibnamefont{Durr}},
  \bibinfo{author}{\bibfnamefont{Z.}~\bibnamefont{Fodor}},
  \bibinfo{author}{\bibfnamefont{S.~D.} \bibnamefont{Katz}},
  \bibinfo{author}{\bibfnamefont{S.}~\bibnamefont{Krieg}}, \bibnamefont{and}
  \bibinfo{author}{\bibfnamefont{K.~K.} \bibnamefont{Szabo}},
  \bibinfo{journal}{JHEP} \textbf{\bibinfo{volume}{06}}, \bibinfo{pages}{088}
  (\bibinfo{year}{2009}), \eprint{0903.4155}.

\bibitem[{\citenamefont{Bazavov et~al.}(2012{\natexlab{a}})}]{Bazavov:2011nk}
\bibinfo{author}{\bibfnamefont{A.}~\bibnamefont{Bazavov}} \bibnamefont{et~al.},
  \bibinfo{journal}{Phys. Rev. D} \textbf{\bibinfo{volume}{85}},
  \bibinfo{pages}{054503} (\bibinfo{year}{2012}{\natexlab{a}}),
  \eprint{1111.1710}.

\bibitem[{\citenamefont{Ejiri}(2008)}]{Ejiri:2008xt}
\bibinfo{author}{\bibfnamefont{S.}~\bibnamefont{Ejiri}},
  \bibinfo{journal}{Phys. Rev. D} \textbf{\bibinfo{volume}{78}},
  \bibinfo{pages}{074507} (\bibinfo{year}{2008}), \eprint{0804.3227}.

\bibitem[{\citenamefont{Karsch and Redlich}(2011)}]{Karsch:2011gg}
\bibinfo{author}{\bibfnamefont{F.}~\bibnamefont{Karsch}} \bibnamefont{and}
  \bibinfo{author}{\bibfnamefont{K.}~\bibnamefont{Redlich}},
  \bibinfo{journal}{Phys. Rev. D} \textbf{\bibinfo{volume}{84}},
  \bibinfo{pages}{051504} (\bibinfo{year}{2011}), \eprint{1107.1412}.

\bibitem[{\citenamefont{Jeon and Koch}(2000)}]{Jeon:2000wg}
\bibinfo{author}{\bibfnamefont{S.}~\bibnamefont{Jeon}} \bibnamefont{and}
  \bibinfo{author}{\bibfnamefont{V.}~\bibnamefont{Koch}},
  \bibinfo{journal}{Phys. Rev. Lett.} \textbf{\bibinfo{volume}{85}},
  \bibinfo{pages}{2076} (\bibinfo{year}{2000}), \eprint{hep-ph/0003168}.

\bibitem[{\citenamefont{Koch}(2010)}]{Koch:2008ia}
\bibinfo{author}{\bibfnamefont{V.}~\bibnamefont{Koch}},
  \emph{\bibinfo{title}{{Hadronic Fluctuations and Correlations}}}
  (\bibinfo{year}{2010}), pp. \bibinfo{pages}{626--652}, \eprint{0810.2520}.

\bibitem[{\citenamefont{Bzdak et~al.}(2018)\citenamefont{Bzdak, Koch,
  Oliinychenko, and Steinheimer}}]{Bzdak:2018uhv}
\bibinfo{author}{\bibfnamefont{A.}~\bibnamefont{Bzdak}},
  \bibinfo{author}{\bibfnamefont{V.}~\bibnamefont{Koch}},
  \bibinfo{author}{\bibfnamefont{D.}~\bibnamefont{Oliinychenko}},
  \bibnamefont{and}
  \bibinfo{author}{\bibfnamefont{J.}~\bibnamefont{Steinheimer}},
  \bibinfo{journal}{Phys. Rev. C} \textbf{\bibinfo{volume}{98}},
  \bibinfo{pages}{054901} (\bibinfo{year}{2018}), \eprint{1804.04463}.

\bibitem[{\citenamefont{Mishustin}(1999)}]{Mishustin:1998eq}
\bibinfo{author}{\bibfnamefont{I.~N.} \bibnamefont{Mishustin}},
  \bibinfo{journal}{Phys. Rev. Lett.} \textbf{\bibinfo{volume}{82}},
  \bibinfo{pages}{4779} (\bibinfo{year}{1999}), \eprint{hep-ph/9811307}.

\bibitem[{\citenamefont{Xu et~al.}(2008)\citenamefont{Xu, Yu, and
  Liu}}]{Xu:2007oam}
\bibinfo{author}{\bibfnamefont{M.}~\bibnamefont{Xu}},
  \bibinfo{author}{\bibfnamefont{M.}~\bibnamefont{Yu}}, \bibnamefont{and}
  \bibinfo{author}{\bibfnamefont{L.}~\bibnamefont{Liu}},
  \bibinfo{journal}{Phys. Rev. Lett.} \textbf{\bibinfo{volume}{100}},
  \bibinfo{pages}{092301} (\bibinfo{year}{2008}), \eprint{0712.1641}.

\bibitem[{\citenamefont{Jin et~al.}(2019)\citenamefont{Jin, Chen, Lin, Ma, Ma,
  and Zhang}}]{Jin:2018fwq}
\bibinfo{author}{\bibfnamefont{X.}~\bibnamefont{Jin}},
  \bibinfo{author}{\bibfnamefont{J.}~\bibnamefont{Chen}},
  \bibinfo{author}{\bibfnamefont{Z.}~\bibnamefont{Lin}},
  \bibinfo{author}{\bibfnamefont{G.}~\bibnamefont{Ma}},
  \bibinfo{author}{\bibfnamefont{Y.}~\bibnamefont{Ma}}, \bibnamefont{and}
  \bibinfo{author}{\bibfnamefont{S.}~\bibnamefont{Zhang}},
  \bibinfo{journal}{Sci. China Phys. Mech. Astron.}
  \textbf{\bibinfo{volume}{62}}, \bibinfo{pages}{11012} (\bibinfo{year}{2019}).

\bibitem[{\citenamefont{Chomaz et~al.}(2004)\citenamefont{Chomaz, Colonna, and
  Randrup}}]{Chomaz:2003dz}
\bibinfo{author}{\bibfnamefont{P.}~\bibnamefont{Chomaz}},
  \bibinfo{author}{\bibfnamefont{M.}~\bibnamefont{Colonna}}, \bibnamefont{and}
  \bibinfo{author}{\bibfnamefont{J.}~\bibnamefont{Randrup}},
  \bibinfo{journal}{Phys. Rept.} \textbf{\bibinfo{volume}{389}},
  \bibinfo{pages}{263} (\bibinfo{year}{2004}).

\bibitem[{\citenamefont{Randrup}(2004)}]{Randrup:2003mu}
\bibinfo{author}{\bibfnamefont{J.}~\bibnamefont{Randrup}},
  \bibinfo{journal}{Phys. Rev. Lett.} \textbf{\bibinfo{volume}{92}},
  \bibinfo{pages}{122301} (\bibinfo{year}{2004}), \eprint{hep-ph/0308271}.

\bibitem[{\citenamefont{Sasaki et~al.}(2007)\citenamefont{Sasaki, Friman, and
  Redlich}}]{Sasaki:2007db}
\bibinfo{author}{\bibfnamefont{C.}~\bibnamefont{Sasaki}},
  \bibinfo{author}{\bibfnamefont{B.}~\bibnamefont{Friman}}, \bibnamefont{and}
  \bibinfo{author}{\bibfnamefont{K.}~\bibnamefont{Redlich}},
  \bibinfo{journal}{Phys. Rev. Lett.} \textbf{\bibinfo{volume}{99}},
  \bibinfo{pages}{232301} (\bibinfo{year}{2007}), \eprint{hep-ph/0702254}.

\bibitem[{\citenamefont{Steinheimer and Randrup}(2012)}]{Steinheimer:2012gc}
\bibinfo{author}{\bibfnamefont{J.}~\bibnamefont{Steinheimer}} \bibnamefont{and}
  \bibinfo{author}{\bibfnamefont{J.}~\bibnamefont{Randrup}},
  \bibinfo{journal}{Phys. Rev. Lett.} \textbf{\bibinfo{volume}{109}},
  \bibinfo{pages}{212301} (\bibinfo{year}{2012}), \eprint{1209.2462}.

\bibitem[{\citenamefont{Wu et~al.}(2021)\citenamefont{Wu, Shen, and
  Song}}]{Wu:2021xgu}
\bibinfo{author}{\bibfnamefont{S.}~\bibnamefont{Wu}},
  \bibinfo{author}{\bibfnamefont{C.}~\bibnamefont{Shen}}, \bibnamefont{and}
  \bibinfo{author}{\bibfnamefont{H.}~\bibnamefont{Song}},
  \bibinfo{journal}{Chin. Phys. Lett.} \textbf{\bibinfo{volume}{38}},
  \bibinfo{pages}{081201} (\bibinfo{year}{2021}), \eprint{2104.13250}.

\bibitem[{\citenamefont{Adams et~al.}(2005)}]{STAR:2005gfr}
\bibinfo{author}{\bibfnamefont{J.}~\bibnamefont{Adams}} \bibnamefont{et~al.}
  (\bibinfo{collaboration}{STAR}), \bibinfo{journal}{Nucl. Phys. A}
  \textbf{\bibinfo{volume}{757}}, \bibinfo{pages}{102} (\bibinfo{year}{2005}),
  \eprint{nucl-ex/0501009}.

\bibitem[{\citenamefont{Aggarwal et~al.}(2010{\natexlab{a}})}]{STAR:2010mib}
\bibinfo{author}{\bibfnamefont{M.~M.} \bibnamefont{Aggarwal}}
  \bibnamefont{et~al.} (\bibinfo{collaboration}{STAR}), \bibinfo{journal}{Phys.
  Rev. Lett.} \textbf{\bibinfo{volume}{105}}, \bibinfo{pages}{022302}
  (\bibinfo{year}{2010}{\natexlab{a}}), \eprint{1004.4959}.

\bibitem[{\citenamefont{Arsene et~al.}(2005)}]{BRAHMS:2004adc}
\bibinfo{author}{\bibfnamefont{I.}~\bibnamefont{Arsene}} \bibnamefont{et~al.}
  (\bibinfo{collaboration}{BRAHMS}), \bibinfo{journal}{Nucl. Phys. A}
  \textbf{\bibinfo{volume}{757}}, \bibinfo{pages}{1} (\bibinfo{year}{2005}),
  \eprint{nucl-ex/0410020}.

\bibitem[{\citenamefont{Friman et~al.}(2011)\citenamefont{Friman, Karsch,
  Redlich, and Skokov}}]{Friman:2011pf}
\bibinfo{author}{\bibfnamefont{B.}~\bibnamefont{Friman}},
  \bibinfo{author}{\bibfnamefont{F.}~\bibnamefont{Karsch}},
  \bibinfo{author}{\bibfnamefont{K.}~\bibnamefont{Redlich}}, \bibnamefont{and}
  \bibinfo{author}{\bibfnamefont{V.}~\bibnamefont{Skokov}},
  \bibinfo{journal}{Eur. Phys. J. C} \textbf{\bibinfo{volume}{71}},
  \bibinfo{pages}{1694} (\bibinfo{year}{2011}), \eprint{1103.3511}.

\bibitem[{\citenamefont{Stephanov}(2009)}]{Stephanov:2008qz}
\bibinfo{author}{\bibfnamefont{M.~A.} \bibnamefont{Stephanov}},
  \bibinfo{journal}{Phys. Rev. Lett.} \textbf{\bibinfo{volume}{102}},
  \bibinfo{pages}{032301} (\bibinfo{year}{2009}), \eprint{0809.3450}.

\bibitem[{\citenamefont{Athanasiou et~al.}(2010)\citenamefont{Athanasiou,
  Rajagopal, and Stephanov}}]{Athanasiou:2010kw}
\bibinfo{author}{\bibfnamefont{C.}~\bibnamefont{Athanasiou}},
  \bibinfo{author}{\bibfnamefont{K.}~\bibnamefont{Rajagopal}},
  \bibnamefont{and}
  \bibinfo{author}{\bibfnamefont{M.}~\bibnamefont{Stephanov}},
  \bibinfo{journal}{Phys. Rev. D} \textbf{\bibinfo{volume}{82}},
  \bibinfo{pages}{074008} (\bibinfo{year}{2010}), \eprint{1006.4636}.

\bibitem[{\citenamefont{Stephanov}(2011)}]{Stephanov:2011pb}
\bibinfo{author}{\bibfnamefont{M.~A.} \bibnamefont{Stephanov}},
  \bibinfo{journal}{Phys. Rev. Lett.} \textbf{\bibinfo{volume}{107}},
  \bibinfo{pages}{052301} (\bibinfo{year}{2011}), \eprint{1104.1627}.

\bibitem[{\citenamefont{Cheng et~al.}(2009)}]{Cheng:2008zh}
\bibinfo{author}{\bibfnamefont{M.}~\bibnamefont{Cheng}} \bibnamefont{et~al.},
  \bibinfo{journal}{Phys. Rev. D} \textbf{\bibinfo{volume}{79}},
  \bibinfo{pages}{074505} (\bibinfo{year}{2009}), \eprint{0811.1006}.

\bibitem[{\citenamefont{Gavai and Gupta}(2011)}]{Gavai:2010zn}
\bibinfo{author}{\bibfnamefont{R.~V.} \bibnamefont{Gavai}} \bibnamefont{and}
  \bibinfo{author}{\bibfnamefont{S.}~\bibnamefont{Gupta}},
  \bibinfo{journal}{Phys. Lett. B} \textbf{\bibinfo{volume}{696}},
  \bibinfo{pages}{459} (\bibinfo{year}{2011}), \eprint{1001.3796}.

\bibitem[{\citenamefont{Chen et~al.}(2021)\citenamefont{Chen, Zhao, Wu, Li, and
  Wu}}]{Chen:2021kjd}
\bibinfo{author}{\bibfnamefont{L.-Z.} \bibnamefont{Chen}},
  \bibinfo{author}{\bibfnamefont{Y.-Y.} \bibnamefont{Zhao}},
  \bibinfo{author}{\bibfnamefont{J.}~\bibnamefont{Wu}},
  \bibinfo{author}{\bibfnamefont{Z.-M.} \bibnamefont{Li}}, \bibnamefont{and}
  \bibinfo{author}{\bibfnamefont{Y.-F.} \bibnamefont{Wu}},
  \bibinfo{journal}{Chin. Phys. C} \textbf{\bibinfo{volume}{45}},
  \bibinfo{pages}{104103} (\bibinfo{year}{2021}), \eprint{2109.14169}.

\bibitem[{\citenamefont{Gupta et~al.}(2011)\citenamefont{Gupta, Luo, Mohanty,
  Ritter, and Xu}}]{Gupta:2011wh}
\bibinfo{author}{\bibfnamefont{S.}~\bibnamefont{Gupta}},
  \bibinfo{author}{\bibfnamefont{X.}~\bibnamefont{Luo}},
  \bibinfo{author}{\bibfnamefont{B.}~\bibnamefont{Mohanty}},
  \bibinfo{author}{\bibfnamefont{H.~G.} \bibnamefont{Ritter}},
  \bibnamefont{and} \bibinfo{author}{\bibfnamefont{N.}~\bibnamefont{Xu}},
  \bibinfo{journal}{Science} \textbf{\bibinfo{volume}{332}},
  \bibinfo{pages}{1525} (\bibinfo{year}{2011}), \eprint{1105.3934}.

\bibitem[{\citenamefont{Bazavov et~al.}(2012{\natexlab{b}})}]{Bazavov:2012vg}
\bibinfo{author}{\bibfnamefont{A.}~\bibnamefont{Bazavov}} \bibnamefont{et~al.},
  \bibinfo{journal}{Phys. Rev. Lett.} \textbf{\bibinfo{volume}{109}},
  \bibinfo{pages}{192302} (\bibinfo{year}{2012}{\natexlab{b}}),
  \eprint{1208.1220}.

\bibitem[{\citenamefont{Aggarwal et~al.}(2010{\natexlab{b}})}]{STAR:2010vob}
\bibinfo{author}{\bibfnamefont{M.~M.} \bibnamefont{Aggarwal}}
  \bibnamefont{et~al.} (\bibinfo{collaboration}{STAR})
  (\bibinfo{year}{2010}{\natexlab{b}}), \eprint{1007.2613}.

\bibitem[{\citenamefont{Adamczyk et~al.}(2017)}]{STAR:2017sal}
\bibinfo{author}{\bibfnamefont{L.}~\bibnamefont{Adamczyk}} \bibnamefont{et~al.}
  (\bibinfo{collaboration}{STAR}), \bibinfo{journal}{Phys. Rev. C}
  \textbf{\bibinfo{volume}{96}}, \bibinfo{pages}{044904}
  (\bibinfo{year}{2017}), \eprint{1701.07065}.

\bibitem[{\citenamefont{Adamczyk et~al.}(2014{\natexlab{a}})}]{STAR:2014egu}
\bibinfo{author}{\bibfnamefont{L.}~\bibnamefont{Adamczyk}} \bibnamefont{et~al.}
  (\bibinfo{collaboration}{STAR}), \bibinfo{journal}{Phys. Rev. Lett.}
  \textbf{\bibinfo{volume}{113}}, \bibinfo{pages}{092301}
  (\bibinfo{year}{2014}{\natexlab{a}}), \eprint{1402.1558}.

\bibitem[{\citenamefont{Adamczyk et~al.}(2014{\natexlab{b}})}]{STAR:2013gus}
\bibinfo{author}{\bibfnamefont{L.}~\bibnamefont{Adamczyk}} \bibnamefont{et~al.}
  (\bibinfo{collaboration}{STAR}), \bibinfo{journal}{Phys. Rev. Lett.}
  \textbf{\bibinfo{volume}{112}}, \bibinfo{pages}{032302}
  (\bibinfo{year}{2014}{\natexlab{b}}), \eprint{1309.5681}.

\bibitem[{\citenamefont{Pandav}(2021)}]{Pandav:2020uzx}
\bibinfo{author}{\bibfnamefont{A.}~\bibnamefont{Pandav}}
  (\bibinfo{collaboration}{STAR}), \bibinfo{journal}{Nucl. Phys. A}
  \textbf{\bibinfo{volume}{1005}}, \bibinfo{pages}{121936}
  (\bibinfo{year}{2021}), \eprint{2003.12503}.

\bibitem[{\citenamefont{Th\"ader}(2016)}]{Thader:2016gpa}
\bibinfo{author}{\bibfnamefont{J.}~\bibnamefont{Th\"ader}}
  (\bibinfo{collaboration}{STAR}), \bibinfo{journal}{Nucl. Phys. A}
  \textbf{\bibinfo{volume}{956}}, \bibinfo{pages}{320} (\bibinfo{year}{2016}),
  \eprint{1601.00951}.

\bibitem[{\citenamefont{Asakawa et~al.}(2000)\citenamefont{Asakawa, Heinz, and
  Muller}}]{Asakawa:2000wh}
\bibinfo{author}{\bibfnamefont{M.}~\bibnamefont{Asakawa}},
  \bibinfo{author}{\bibfnamefont{U.~W.} \bibnamefont{Heinz}}, \bibnamefont{and}
  \bibinfo{author}{\bibfnamefont{B.}~\bibnamefont{Muller}},
  \bibinfo{journal}{Phys. Rev. Lett.} \textbf{\bibinfo{volume}{85}},
  \bibinfo{pages}{2072} (\bibinfo{year}{2000}), \eprint{hep-ph/0003169}.

\bibitem[{\citenamefont{Asakawa et~al.}(2009)\citenamefont{Asakawa, Ejiri, and
  Kitazawa}}]{Asakawa:2009aj}
\bibinfo{author}{\bibfnamefont{M.}~\bibnamefont{Asakawa}},
  \bibinfo{author}{\bibfnamefont{S.}~\bibnamefont{Ejiri}}, \bibnamefont{and}
  \bibinfo{author}{\bibfnamefont{M.}~\bibnamefont{Kitazawa}},
  \bibinfo{journal}{Phys. Rev. Lett.} \textbf{\bibinfo{volume}{103}},
  \bibinfo{pages}{262301} (\bibinfo{year}{2009}), \eprint{0904.2089}.

\bibitem[{\citenamefont{Abdallah et~al.}(2021)}]{STAR:2021iop}
\bibinfo{author}{\bibfnamefont{M.}~\bibnamefont{Abdallah}} \bibnamefont{et~al.}
  (\bibinfo{collaboration}{STAR}), \bibinfo{journal}{Phys. Rev. C}
  \textbf{\bibinfo{volume}{104}}, \bibinfo{pages}{024902}
  (\bibinfo{year}{2021}), \eprint{2101.12413}.

\bibitem[{\citenamefont{Bzdak et~al.}(2017{\natexlab{a}})\citenamefont{Bzdak,
  Koch, and Strodthoff}}]{Bzdak:2016sxg}
\bibinfo{author}{\bibfnamefont{A.}~\bibnamefont{Bzdak}},
  \bibinfo{author}{\bibfnamefont{V.}~\bibnamefont{Koch}}, \bibnamefont{and}
  \bibinfo{author}{\bibfnamefont{N.}~\bibnamefont{Strodthoff}},
  \bibinfo{journal}{Phys. Rev. C} \textbf{\bibinfo{volume}{95}},
  \bibinfo{pages}{054906} (\bibinfo{year}{2017}{\natexlab{a}}),
  \eprint{1607.07375}.

\bibitem[{\citenamefont{Ling and Stephanov}(2016)}]{Ling:2015yau}
\bibinfo{author}{\bibfnamefont{B.}~\bibnamefont{Ling}} \bibnamefont{and}
  \bibinfo{author}{\bibfnamefont{M.~A.} \bibnamefont{Stephanov}},
  \bibinfo{journal}{Phys. Rev. C} \textbf{\bibinfo{volume}{93}},
  \bibinfo{pages}{034915} (\bibinfo{year}{2016}), \eprint{1512.09125}.

\bibitem[{\citenamefont{Fu et~al.}(2021)\citenamefont{Fu, Luo, Pawlowski,
  Rennecke, Wen, and Yin}}]{Fu:2021oaw}
\bibinfo{author}{\bibfnamefont{W.-j.} \bibnamefont{Fu}},
  \bibinfo{author}{\bibfnamefont{X.}~\bibnamefont{Luo}},
  \bibinfo{author}{\bibfnamefont{J.~M.} \bibnamefont{Pawlowski}},
  \bibinfo{author}{\bibfnamefont{F.}~\bibnamefont{Rennecke}},
  \bibinfo{author}{\bibfnamefont{R.}~\bibnamefont{Wen}}, \bibnamefont{and}
  \bibinfo{author}{\bibfnamefont{S.}~\bibnamefont{Yin}},
  \bibinfo{journal}{Phys. Rev. D} \textbf{\bibinfo{volume}{104}},
  \bibinfo{pages}{094047} (\bibinfo{year}{2021}), \eprint{2101.06035}.

\bibitem[{\citenamefont{Vovchenko et~al.}(2022)\citenamefont{Vovchenko, Koch,
  and Shen}}]{Vovchenko:2021kxx}
\bibinfo{author}{\bibfnamefont{V.}~\bibnamefont{Vovchenko}},
  \bibinfo{author}{\bibfnamefont{V.}~\bibnamefont{Koch}}, \bibnamefont{and}
  \bibinfo{author}{\bibfnamefont{C.}~\bibnamefont{Shen}},
  \bibinfo{journal}{Phys. Rev. C} \textbf{\bibinfo{volume}{105}},
  \bibinfo{pages}{014904} (\bibinfo{year}{2022}), \eprint{2107.00163}.

\bibitem[{\citenamefont{Lin et~al.}(2005)\citenamefont{Lin, Ko, Li, Zhang, and
  Pal}}]{Lin:2004en}
\bibinfo{author}{\bibfnamefont{Z.-W.} \bibnamefont{Lin}},
  \bibinfo{author}{\bibfnamefont{C.~M.} \bibnamefont{Ko}},
  \bibinfo{author}{\bibfnamefont{B.-A.} \bibnamefont{Li}},
  \bibinfo{author}{\bibfnamefont{B.}~\bibnamefont{Zhang}}, \bibnamefont{and}
  \bibinfo{author}{\bibfnamefont{S.}~\bibnamefont{Pal}},
  \bibinfo{journal}{Phys. Rev. C} \textbf{\bibinfo{volume}{72}},
  \bibinfo{pages}{064901} (\bibinfo{year}{2005}), \eprint{nucl-th/0411110}.

\bibitem[{\citenamefont{Lin and Zheng}(2021)}]{Lin:2021mdn}
\bibinfo{author}{\bibfnamefont{Z.-W.} \bibnamefont{Lin}} \bibnamefont{and}
  \bibinfo{author}{\bibfnamefont{L.}~\bibnamefont{Zheng}},
  \bibinfo{journal}{Nucl. Sci. Tech.} \textbf{\bibinfo{volume}{32}},
  \bibinfo{pages}{113} (\bibinfo{year}{2021}), \eprint{2110.02989}.

\bibitem[{\citenamefont{Ma and Lin}(2016)}]{Ma:2016fve}
\bibinfo{author}{\bibfnamefont{G.-L.} \bibnamefont{Ma}} \bibnamefont{and}
  \bibinfo{author}{\bibfnamefont{Z.-W.} \bibnamefont{Lin}},
  \bibinfo{journal}{Phys. Rev. C} \textbf{\bibinfo{volume}{93}},
  \bibinfo{pages}{054911} (\bibinfo{year}{2016}), \eprint{1601.08160}.

\bibitem[{\citenamefont{Ma and Zhang}(2011)}]{Ma:2011uma}
\bibinfo{author}{\bibfnamefont{G.-L.} \bibnamefont{Ma}} \bibnamefont{and}
  \bibinfo{author}{\bibfnamefont{B.}~\bibnamefont{Zhang}},
  \bibinfo{journal}{Phys. Lett. B} \textbf{\bibinfo{volume}{700}},
  \bibinfo{pages}{39} (\bibinfo{year}{2011}), \eprint{1101.1701}.

\bibitem[{\citenamefont{Ma}(2013)}]{Ma:2013pha}
\bibinfo{author}{\bibfnamefont{G.-L.} \bibnamefont{Ma}},
  \bibinfo{journal}{Phys. Rev. C} \textbf{\bibinfo{volume}{87}},
  \bibinfo{pages}{064901} (\bibinfo{year}{2013}), \eprint{1304.2841}.

\bibitem[{\citenamefont{Bozek et~al.}(2015)\citenamefont{Bozek, Bzdak, and
  Ma}}]{Bozek:2015swa}
\bibinfo{author}{\bibfnamefont{P.}~\bibnamefont{Bozek}},
  \bibinfo{author}{\bibfnamefont{A.}~\bibnamefont{Bzdak}}, \bibnamefont{and}
  \bibinfo{author}{\bibfnamefont{G.-L.} \bibnamefont{Ma}},
  \bibinfo{journal}{Phys. Lett. B} \textbf{\bibinfo{volume}{748}},
  \bibinfo{pages}{301} (\bibinfo{year}{2015}), \eprint{1503.03655}.

\bibitem[{\citenamefont{Bzdak and Ma}(2014)}]{Bzdak:2014dia}
\bibinfo{author}{\bibfnamefont{A.}~\bibnamefont{Bzdak}} \bibnamefont{and}
  \bibinfo{author}{\bibfnamefont{G.-L.} \bibnamefont{Ma}},
  \bibinfo{journal}{Phys. Rev. Lett.} \textbf{\bibinfo{volume}{113}},
  \bibinfo{pages}{252301} (\bibinfo{year}{2014}), \eprint{1406.2804}.

\bibitem[{\citenamefont{Wang and Gyulassy}(1991)}]{Wang:1991hta}
\bibinfo{author}{\bibfnamefont{X.-N.} \bibnamefont{Wang}} \bibnamefont{and}
  \bibinfo{author}{\bibfnamefont{M.}~\bibnamefont{Gyulassy}},
  \bibinfo{journal}{Phys. Rev. D} \textbf{\bibinfo{volume}{44}},
  \bibinfo{pages}{3501} (\bibinfo{year}{1991}).

\bibitem[{\citenamefont{Gyulassy and Wang}(1994)}]{Gyulassy:1994ew}
\bibinfo{author}{\bibfnamefont{M.}~\bibnamefont{Gyulassy}} \bibnamefont{and}
  \bibinfo{author}{\bibfnamefont{X.-N.} \bibnamefont{Wang}},
  \bibinfo{journal}{Comput. Phys. Commun.} \textbf{\bibinfo{volume}{83}},
  \bibinfo{pages}{307} (\bibinfo{year}{1994}), \eprint{nucl-th/9502021}.

\bibitem[{\citenamefont{Zhang}(1998)}]{Zhang:1997ej}
\bibinfo{author}{\bibfnamefont{B.}~\bibnamefont{Zhang}},
  \bibinfo{journal}{Comput. Phys. Commun.} \textbf{\bibinfo{volume}{109}},
  \bibinfo{pages}{193} (\bibinfo{year}{1998}), \eprint{nucl-th/9709009}.

\bibitem[{\citenamefont{Li and Ko}(1995)}]{Li:1995pra}
\bibinfo{author}{\bibfnamefont{B.-A.} \bibnamefont{Li}} \bibnamefont{and}
  \bibinfo{author}{\bibfnamefont{C.~M.} \bibnamefont{Ko}},
  \bibinfo{journal}{Phys. Rev. C} \textbf{\bibinfo{volume}{52}},
  \bibinfo{pages}{2037} (\bibinfo{year}{1995}), \eprint{nucl-th/9505016}.

\bibitem[{\citenamefont{Lin}(2014)}]{Lin:2014uwa}
\bibinfo{author}{\bibfnamefont{Z.~W.} \bibnamefont{Lin}},
  \bibinfo{journal}{Acta Phys. Polon. Supp.} \textbf{\bibinfo{volume}{7}},
  \bibinfo{pages}{191} (\bibinfo{year}{2014}), \eprint{1403.1854}.

\bibitem[{\citenamefont{Huang and Ma}(2021)}]{Huang:2021ihy}
\bibinfo{author}{\bibfnamefont{L.}~\bibnamefont{Huang}} \bibnamefont{and}
  \bibinfo{author}{\bibfnamefont{G.-L.} \bibnamefont{Ma}},
  \bibinfo{journal}{Chin. Phys. C} \textbf{\bibinfo{volume}{45}},
  \bibinfo{pages}{074110} (\bibinfo{year}{2021}), \eprint{2107.09264}.

\bibitem[{\citenamefont{Kitazawa and
  Asakawa}(2012{\natexlab{a}})}]{Kitazawa:2011wh}
\bibinfo{author}{\bibfnamefont{M.}~\bibnamefont{Kitazawa}} \bibnamefont{and}
  \bibinfo{author}{\bibfnamefont{M.}~\bibnamefont{Asakawa}},
  \bibinfo{journal}{Phys. Rev. C} \textbf{\bibinfo{volume}{85}},
  \bibinfo{pages}{021901} (\bibinfo{year}{2012}{\natexlab{a}}),
  \eprint{1107.2755}.

\bibitem[{\citenamefont{Kitazawa and
  Asakawa}(2012{\natexlab{b}})}]{Kitazawa:2012at}
\bibinfo{author}{\bibfnamefont{M.}~\bibnamefont{Kitazawa}} \bibnamefont{and}
  \bibinfo{author}{\bibfnamefont{M.}~\bibnamefont{Asakawa}},
  \bibinfo{journal}{Phys. Rev. C} \textbf{\bibinfo{volume}{86}},
  \bibinfo{pages}{024904} (\bibinfo{year}{2012}{\natexlab{b}}),
  \bibinfo{note}{[Erratum: Phys.Rev.C 86, 069902 (2012)]}, \eprint{1205.3292}.

\bibitem[{\citenamefont{Luo et~al.}(2012)\citenamefont{Luo, Mohanty, Ritter,
  and Xu}}]{Luo:2011rg}
\bibinfo{author}{\bibfnamefont{X.-F.} \bibnamefont{Luo}},
  \bibinfo{author}{\bibfnamefont{B.}~\bibnamefont{Mohanty}},
  \bibinfo{author}{\bibfnamefont{H.~G.} \bibnamefont{Ritter}},
  \bibnamefont{and} \bibinfo{author}{\bibfnamefont{N.}~\bibnamefont{Xu}},
  \bibinfo{journal}{Phys. Atom. Nucl.} \textbf{\bibinfo{volume}{75}},
  \bibinfo{pages}{676} (\bibinfo{year}{2012}), \eprint{1105.5049}.

\bibitem[{\citenamefont{Luo et~al.}(2010)\citenamefont{Luo, Mohanty, Ritter,
  and Xu}}]{Luo:2010by}
\bibinfo{author}{\bibfnamefont{X.~F.} \bibnamefont{Luo}},
  \bibinfo{author}{\bibfnamefont{B.}~\bibnamefont{Mohanty}},
  \bibinfo{author}{\bibfnamefont{H.~G.} \bibnamefont{Ritter}},
  \bibnamefont{and} \bibinfo{author}{\bibfnamefont{N.}~\bibnamefont{Xu}},
  \bibinfo{journal}{J. Phys. G} \textbf{\bibinfo{volume}{37}},
  \bibinfo{pages}{094061} (\bibinfo{year}{2010}), \eprint{1001.2847}.

\bibitem[{\citenamefont{Luo}(2011)}]{Luo:2011ts}
\bibinfo{author}{\bibfnamefont{X.-F.} \bibnamefont{Luo}}
  (\bibinfo{collaboration}{STAR}), \bibinfo{journal}{J. Phys. Conf. Ser.}
  \textbf{\bibinfo{volume}{316}}, \bibinfo{pages}{012003}
  (\bibinfo{year}{2011}), \eprint{1106.2926}.

\bibitem[{\citenamefont{Luo and Xu}(2017)}]{Luo:2017faz}
\bibinfo{author}{\bibfnamefont{X.}~\bibnamefont{Luo}} \bibnamefont{and}
  \bibinfo{author}{\bibfnamefont{N.}~\bibnamefont{Xu}}, \bibinfo{journal}{Nucl.
  Sci. Tech.} \textbf{\bibinfo{volume}{28}}, \bibinfo{pages}{112}
  (\bibinfo{year}{2017}), \eprint{1701.02105}.

\bibitem[{\citenamefont{Kitazawa and Luo}(2017)}]{Kitazawa:2017ljq}
\bibinfo{author}{\bibfnamefont{M.}~\bibnamefont{Kitazawa}} \bibnamefont{and}
  \bibinfo{author}{\bibfnamefont{X.}~\bibnamefont{Luo}},
  \bibinfo{journal}{Phys. Rev. C} \textbf{\bibinfo{volume}{96}},
  \bibinfo{pages}{024910} (\bibinfo{year}{2017}), \eprint{1704.04909}.

\bibitem[{\citenamefont{Adam et~al.}(2021)}]{STAR:2020tga}
\bibinfo{author}{\bibfnamefont{J.}~\bibnamefont{Adam}} \bibnamefont{et~al.}
  (\bibinfo{collaboration}{STAR}), \bibinfo{journal}{Phys. Rev. Lett.}
  \textbf{\bibinfo{volume}{126}}, \bibinfo{pages}{092301}
  (\bibinfo{year}{2021}), \eprint{2001.02852}.

\bibitem[{\citenamefont{Bzdak et~al.}(2017{\natexlab{b}})\citenamefont{Bzdak,
  Koch, and Skokov}}]{Bzdak:2016jxo}
\bibinfo{author}{\bibfnamefont{A.}~\bibnamefont{Bzdak}},
  \bibinfo{author}{\bibfnamefont{V.}~\bibnamefont{Koch}}, \bibnamefont{and}
  \bibinfo{author}{\bibfnamefont{V.}~\bibnamefont{Skokov}},
  \bibinfo{journal}{Eur. Phys. J. C} \textbf{\bibinfo{volume}{77}},
  \bibinfo{pages}{288} (\bibinfo{year}{2017}{\natexlab{b}}),
  \eprint{1612.05128}.

\bibitem[{\citenamefont{Luo et~al.}(2013)\citenamefont{Luo, Xu, Mohanty, and
  Xu}}]{Luo:2013bmi}
\bibinfo{author}{\bibfnamefont{X.}~\bibnamefont{Luo}},
  \bibinfo{author}{\bibfnamefont{J.}~\bibnamefont{Xu}},
  \bibinfo{author}{\bibfnamefont{B.}~\bibnamefont{Mohanty}}, \bibnamefont{and}
  \bibinfo{author}{\bibfnamefont{N.}~\bibnamefont{Xu}}, \bibinfo{journal}{J.
  Phys. G} \textbf{\bibinfo{volume}{40}}, \bibinfo{pages}{105104}
  (\bibinfo{year}{2013}), \eprint{1302.2332}.

\bibitem[{\citenamefont{He and Luo}(2018)}]{He:2018mri}
\bibinfo{author}{\bibfnamefont{S.}~\bibnamefont{He}} \bibnamefont{and}
  \bibinfo{author}{\bibfnamefont{X.}~\bibnamefont{Luo}},
  \bibinfo{journal}{Chin. Phys. C} \textbf{\bibinfo{volume}{42}},
  \bibinfo{pages}{104001} (\bibinfo{year}{2018}), \eprint{1802.02911}.

\bibitem[{\citenamefont{Bazavov et~al.}(2017)}]{Bazavov:2017dus}
\bibinfo{author}{\bibfnamefont{A.}~\bibnamefont{Bazavov}} \bibnamefont{et~al.},
  \bibinfo{journal}{Phys. Rev. D} \textbf{\bibinfo{volume}{95}},
  \bibinfo{pages}{054504} (\bibinfo{year}{2017}), \eprint{1701.04325}.

\bibitem[{\citenamefont{Lin et~al.}(2017)\citenamefont{Lin, Chen, and
  Li}}]{Lin:2017xkd}
\bibinfo{author}{\bibfnamefont{Y.}~\bibnamefont{Lin}},
  \bibinfo{author}{\bibfnamefont{L.}~\bibnamefont{Chen}}, \bibnamefont{and}
  \bibinfo{author}{\bibfnamefont{Z.}~\bibnamefont{Li}}, \bibinfo{journal}{Phys.
  Rev. C} \textbf{\bibinfo{volume}{96}}, \bibinfo{pages}{044906}
  (\bibinfo{year}{2017}), \eprint{1707.04375}.

\bibitem[{\citenamefont{He and Lin}(2017)}]{He:2017tla}
\bibinfo{author}{\bibfnamefont{Y.}~\bibnamefont{He}} \bibnamefont{and}
  \bibinfo{author}{\bibfnamefont{Z.-W.} \bibnamefont{Lin}},
  \bibinfo{journal}{Phys. Rev. C} \textbf{\bibinfo{volume}{96}},
  \bibinfo{pages}{014910} (\bibinfo{year}{2017}), \eprint{1703.02673}.

\bibitem[{\citenamefont{Bzdak and Koch}(2017)}]{Bzdak:2017ltv}
\bibinfo{author}{\bibfnamefont{A.}~\bibnamefont{Bzdak}} \bibnamefont{and}
  \bibinfo{author}{\bibfnamefont{V.}~\bibnamefont{Koch}},
  \bibinfo{journal}{Phys. Rev. C} \textbf{\bibinfo{volume}{96}},
  \bibinfo{pages}{054905} (\bibinfo{year}{2017}), \eprint{1707.02640}.

\bibitem[{\citenamefont{He and Luo}(2017)}]{He:2017zpg}
\bibinfo{author}{\bibfnamefont{S.}~\bibnamefont{He}} \bibnamefont{and}
  \bibinfo{author}{\bibfnamefont{X.}~\bibnamefont{Luo}},
  \bibinfo{journal}{Phys. Lett. B} \textbf{\bibinfo{volume}{774}},
  \bibinfo{pages}{623} (\bibinfo{year}{2017}), \eprint{1704.00423}.

\bibitem[{\citenamefont{Bzdak and Koch}(2012)}]{Bzdak:2012ab}
\bibinfo{author}{\bibfnamefont{A.}~\bibnamefont{Bzdak}} \bibnamefont{and}
  \bibinfo{author}{\bibfnamefont{V.}~\bibnamefont{Koch}},
  \bibinfo{journal}{Phys. Rev. C} \textbf{\bibinfo{volume}{86}},
  \bibinfo{pages}{044904} (\bibinfo{year}{2012}), \eprint{1206.4286}.

\end{thebibliography}


\end{CJK*}

\end{document}